\documentclass[acmsmall]{acmart}
\usepackage{makecell} 
\AtBeginDocument{%
  }

\setcopyright{cc}

\setcctype{by}

\acmJournal{PACMHCI}

\acmYear{2026} \acmVolume{10} \acmNumber{6} \acmArticle{CSCW124}

\acmMonth{10} \acmDOI{10.1145/3816972}

\begin{document}

%%
%% The "title" command has an optional parameter,
%% allowing the author to define a "short title" to be used in page headers.
\title{Mind the Trust Gap: Identifying (Mis)alignments in Teacher-Student Views Toward Control and Agency in K-12 Classroom AI}

%%
%% The "author" command and its associated commands are used to define
%% the authors and their affiliations.
%% Of note is the shared affiliation of the first two authors, and the
%% "authornote" and "authornotemark" commands
%% used to denote shared contribution to the research.
\author{Tomohiro Nagashima}
\email{nagashima@cs.uni-saarland.de}
\orcid{0000-0003-2489-5016}
\affiliation{%
  \institution{Saarland University, Saarland Informatics Campus}
  \city{Saarbrücken}
  \country{Germany}
}

\author{Lisa Siegrist}
\orcid{0009-0009-2484-2214}
\affiliation{%
  \institution{University of St. Gallen}
  \city{St. Gallen}
  \country{Switzerland}}
\email{lisa.siegrist@unisgh.ch}

\author{Niklas Scholz}
\orcid{0009-0007-5754-6011}
\affiliation{%
  \institution{RWTH Aachen University}
  \city{Aachen}
  \country{Germany}}                 
\email{niklas.scholz@rwth-aachen.de}   

\author{Shintaro Sato}
\orcid{0009-0002-8790-9411}
\affiliation{%
  \institution{Saarland University, Saarland Informatics Campus}
  \city{Saarbrücken}
  \country{Germany}}
 \email{sato@cs.uni-saarland.de}

\author{Martina Vincoli}
\orcid{0009-0004-0706-9470}
\affiliation{%
  \institution{ETH Zürich}
  \city{Zürich}
  \country{Switzerland}}
   \email{martina.vincoli@gess.ethz.ch}

\author{Man Su}
\orcid{0000-0001-5696-8414}
\affiliation{%
  \institution{The Leibniz-Institut für Wissensmedien (IWM)}
  \city{Tübingen}
  \country{Germany}}
\email{m.su@iwm-tuebingen.de}

%%
%% By default, the full list of authors will be used in the page
%% headers. Often, this list is too long, and will overlap
%% other information printed in the page headers. This command allows
%% the author to define a more concise list
%% of authors' names for this purpose.
\renewcommand{\shortauthors}{Nagashima et al.}

%%
%% The abstract is a short summary of the work to be presented in the
%% article.
\begin{abstract}
As Artificial Intelligence (AI)-based technologies have been integrated into school classrooms where multiple stakeholders (with different roles) interact with each other, it is critical to deeply understand stakeholder views in the classroom. In particular, prior work has not fully uncovered how teachers’ and school students’ views might or might not align well with each other, especially in K-12 classrooms. We conducted a speed-dating study using storyboards with 16 school students and 15 school teachers in Germany to investigate alignments and misalignments between their views on student-AI decision-making control in K-12 classroom. Through an explicit \textit{pair-matching} analysis, we found that students and teachers had misaligned views on several key topics, including how much they trust AI and social and emotional aspects of student learning with AI. Findings also revealed the importance of teacher-student relationships outside of AI use that shape stakeholders’ views and interactions. We discuss potential reasons for the observed misaligned views and strategies to fill the perspective gaps. This study illustrates the complexities of preferences in teacher-student-AI interactions that depend on the dynamic relations among the stakeholders.
\end{abstract}

%%
%% The code below is generated by the tool at http://dl.acm.org/ccs.cfm.
%% Please copy and paste the code instead of the example below.
%%


%%
%% Keywords. The author(s) should pick words that accurately describe
%% the work being presented. Separate the keywords with commas.

\received{May 13, 2025}
\received[revised]{January 13, 2026}
\received[accepted]{March 17, 2026}

%%
%% This command processes the author and affiliation and title
%% information and builds the first part of the formatted document.
\maketitle

\section{Introduction}
Recent years have witnessed an increased integration of Artificial Intelligence (AI)-based technologies into everyday environments where multiple stakeholders interact with each other in a complex manner (e.g., voice assistance for family home) \cite{garg2022last,lovato2019hey,nagashima2025understanding}. Of many everyday environments, one of the important, complex places where multiple stakeholders interact with each other with AI is K-12 school classrooms \cite{nagashima2025understanding}. Many school classrooms have integrated AI-embedded educational technology to make learning effective and efficient \cite{ahmad2021artificial, chen2020artificial}. For instance, U.S. schools have been using intelligent tutors to adaptively support student learning \cite{pane2014effectiveness, 10.1145/2876034.2876039}. These technologies assist individual students through adaptive scaffolding, personalized and immediate feedback, and content assignment using AI-based, real-time detection of students’ knowledge and affective states \cite{guo2021evolution, koedinger2007exploring, aleven2016instruction, kulik2016effectiveness, long2014gamification, d2013selective}.

When AI-based learning software is used in the classroom, interactions among the AI system, students, and teachers can become highly complex \cite{xu2022systematic}. For example, when a class of students are working with intelligent tutoring software on their own device, students might be assigned content that is not aligned with the curriculum/teacher’s plan, or the teacher may ask all students to stop working on the tutor to explain common mistakes that the teacher noticed \cite{holstein2020conceptual}. In such situations, students may want to request the AI system to assign different tasks, teachers may want to automatically halt students’ progress to get the attention of the whole class (which often has 25-40 students) to explain misconceptions, or teachers may want to pair students up so that well-performing students could help struggling students in solving problems provided by the AI-based tutoring system (which might effectively reduce teacher workload while benefiting students) \cite{echeverria2023designing, holstein2020conceptual,yang2023pair}. Within the CSCW and CSCL (Computer-Supported Collaborative Learning) communities, studies have well investigated these complex, dynamic interactions among stakeholders \cite{echeverria2023designing, 10.1145/3711104, lawrence2024teachers, lawrence2022co, yang2021surveying}, highlighting the critical role that contextual and social factors play in enabling effective interactions with advanced, AI-based systems within a unique classroom environment.

Despite the known complexity of classroom interactions with digital technologies and AI-based tutoring systems in K-12 education \cite{dillenbourg2010technology, echeverria2023designing, feng2023effectiveness, yang2023pair} and the importance of understanding stakeholder views in CSCW, HCI, and AI in Education communities \cite{alfredo2024human, holstein2019co, 10.1145/3711075, zhang2023deliberating}, prior work in this domain has primarily focused on considering or understanding K-12 teachers’ views alone \cite{yang2021surveying, molenaar2022concept, molenaar2022towards, lawrence2022co, lawrence2024teachers, holstein2017intelligent}. With the rise of Generative AI tools (e.g., ChatGPT), more recent studies in CSCW/HCI have examined university students’ perceptions of AI use \cite{adnin2025examining, amoozadeh2024trust, barrett2023not, gorichanaz2023accused, 10.1145/3711104, han2026ai, luo2025does, 10.1145/3637358, zastudil2023generative, wu2024reacting}, with some work surveying both instructors’ and university students’ views toward Generative AI \cite{adnin2025examining, barrett2023not, han2024teachers, wu2024reacting, zastudil2023generative}. However, the university context differs substantially from K-12 classrooms when it comes to understanding stakeholder perspectives. For instance, university students are generally more metacognitively developed than adolescents and are therefore expected to be able to assess AI capabilities appropriately \cite{10.1145/3711104} and to reason about how other stakeholders may feel and behave \cite{adnin2025examining, 10.1145/3711104}. Prior research has also argued that the formation of educator-learner trust differs between K-12 classrooms and higher education settings due to the different ways of attributing authority to educators \cite{kovavc2010trusting}. Moreover, in K-12 education, the use of Generative AI largely occurs outside the context of complex classroom interactions (e.g., primarily for lesson planning and assessment development \cite{cheah2025integrating, duan2024effects}). Within K-12 education, unlike higher education \cite{mohebi2024empowering}, AI-based tutoring systems remain a dominant form of AI technology actively used for teaching and learning, and have been studied in research \cite{pane2014effectiveness, yang2023pair}. As a result, Generative AI is a less central target for examining complex stakeholder views and interactions in K-12 classroom contexts today.

Even when K-12 students’ and teachers’ views are considered in few studies \cite{holstein2019designing, echeverria2023designing}, these studies show students’ and teachers’ preferences rather at a surface level (without comparing/contrasting views between teachers and students at the deeper level of data structure), and do not encompass multiple key aspects of AI use in the classroom that are critical in today’s complex classrooms, including automatic assignment of learning content, data sharing, help-seeking/help-giving, and human-AI co-orchestration, as highlighted by \cite{vincoli2025multidimensional}. We argue that a comprehensive, deeper understanding of how teachers’ and students’ perspectives align or misalign is critical for designing AI systems that foster effective and sustainable classroom interactions among students, teachers, and AI systems where all stakeholders are comfortable with how AI systems are used.

In this study, we ask the following research question: \textit{What preferences do school students and teachers in Germany have regarding student versus AI decision-making control in AI-based learning systems in the classroom, and how do these preferences (mis)align?} We conducted a speed-dating study using storyboards with 16 school students and 15 teachers in Germany. Speed dating is a design method where designers present participants possible futuristic scenarios to get reactions and reflections on what they would accept (or would not accept) and how they would imagine themselves behaving in the scenarios \cite{zimmerman2017speed}. We analyzed video data through Affinity Diagramming \cite{holtzblatt2022consolidation, krause2024affinity}, which were then carefully matched between teacher and student data through a \textit{pair-matching} analysis. Results showed that students and teachers, while sharing some similar perspectives on AI’s decision making in the classroom, also had distinctly different views. In particular, results highlight perspective gaps in the areas of trust in AI, social and emotional aspects of student learning, and data sharing through AI systems. Further, we found that how students and teachers would want to interact with AI systems in the classroom largely depends on what kinds of relationships exist between teachers and students, outside of AI use, which prior work has not fully revealed.

Our work offers three main contributions to CSCW and broader HCI and AI in Education communities. First, we (1) demonstrate the first deep understanding of K-12 students’ and teachers’ views on how they prefer to (and not to) use AI tools in the complex classroom environment, gathered on multiple key aspects of AI use that are relevant in today’s AI-supported classrooms  \cite{vincoli2025multidimensional}. Further, (2) our fine-grained analysis of qualitative data allowed us to contribute an in-depth understanding of alignments and misalignments among their views, which past work has not captured in depth \cite{holstein2019designing, echeverria2023designing}. Among others, our findings illustrate dynamic sociotechnical trade-offs in using AI systems in the classroom, where human-human relationships (between teachers and school students) and the contextual factors intersect with technical affordances of AI systems to realize effective use of AI in the classroom, considering both stakeholder views. Finally, (3) to overcome past work which offers limited actionable insights, particularly for K-12 education stakeholders, we discuss where the misalignments might arise and how the misalignments can be addressed.

\section{Related Work}
\subsection{AI in School Classrooms: Towards Hybrid Intelligence}

AI use in the classroom often translates into the development and evaluation of adaptive learning systems. These technologies, such as Intelligent Tutoring Systems (ITSs) \cite{aleven2016instruction, guo2021evolution, kulik2016effectiveness, pane2014effectiveness}, track students’ learning progress, problem-solving status and predicted knowledge/skill state to adapt their content, feedback, and hints to individual students \cite{aleven2016help, nagashima2023promoting, d2014confusion}. These AI-based technologies have been successfully used in actual school classrooms to enhance student learning, especially in North America \cite{pane2014effectiveness, 10.1145/2876034.2876039}. However, integrating such technologies can create complex interactions in the classroom. Studies have documented the importance of how teachers and students would actively interact with AI technologies to shape how the AI technologies should be used in the classroom environment \cite{chen2022human, dillenbourg2018classroom, holstein2020conceptual, kessler2020exploring, molenaar2022concept}. To conceptualize such a dynamic interplay between AI and human actors, Molenaar \cite{molenaar2022concept} proposed a framework that illustrates to what degree teachers could offload their task vs. how AI could take over teachers’ tasks through phases of teaching automation that could happen (e.g., “Partial Automation” where AI takes care of specific tasks while the teacher is there to monitor the technology). Following such a “hybrid intelligence” view between human teachers and AI tutors \cite{karumbaiah2023spatiotemporal, lawrence2024teachers, lawrence2022co}, researchers have also co-designed AI-based technologies with teachers to integrate teachers’ preferences into the design \cite{holstein2019co, lawrence2024teachers, nazaretsky2022empowering}. These studies typically reveal that teachers do not wish to let AI handle all the tasks they would normally take \cite{echeverria2023designing, lawrence2024teachers, lawrence2022co}. Teacher’s active intervention with the use of AI has also empirically been tested and proven effective, where teachers equipped with AI-based analytics of students’ learning progress within an ITS better supported student learning \cite{holstein2018student}.

Still, this line of work on “hybrid intelligence” \cite{holstein2020conceptual, molenaar2022concept, molenaar2022towards} that combines human intelligence and artificial intelligence in the classroom setting mostly focuses on teacher-AI interactions, often leaving out K-12 students’ perspectives \cite{holstein2017intelligent, lawrence2022co, yang2021surveying}. A large number of works within HCI around teenagers and children with AI happens outside of the classroom environment where they would encounter AI-based systems \cite{10.1145/3687035}. For example, studies have investigated how children perceive conversational agents (e.g., Alexa) and how these agents change the ways in which children communicate \cite{williams2019artificial, lovato2019hey, druga2017hey}, where some found that children do not have a correct understanding of AI \cite{garg2022last}. Parents can influence how children interact with AI systems  \cite{10.1145/3381002} but these interactions between children, AI, and parents’ involvement would not transfer to learning environments (e.g., classroom). Interactions in the classroom tend to be more complex and dynamic compared to situations such as using conversational agents at home; in the classroom (with 25-40 students), even when students interact with an AI-based system individually, teachers may intervene to get the attention of the whole class, may walk to students to offer help, or students might work together with their peers to collaboratively solve problems \cite{echeverria2023designing, yang2023pair, yang2021surveying}. 

\subsection{Student-AI Decision-Making Control when Learning with AI}

A central concern about the AI use in the classroom is how much control different stakeholders would have when making decisions in the classroom \cite{vincoli2025multidimensional, holstein2020conceptual, brod2023agency}. Decision making in the classroom has been a focus of education research before the use of any AI technology to foster students’ autonomous learning behaviors \cite{snow2015does}. However, with the use of AI tools—another active decision-making agent in the classroom—the decision-making space becomes increasingly more complex \cite{molenaar2022concept}. For example, when deciding what content to learn in AI-based software, some systems allow students to select content based on what the students want to practice so that the systems can foster/respect students’ self-regulated skills \cite{long2014gamification, sawyer2017more}. On the other hand, a system may choose the \textit{best} task to assign based on its understanding of what each student needs to practice further \cite{aleven2016instruction}. Or, teachers may intervene in this student-AI interaction, assigning customized content based on the teachers’ understanding of what students would benefit from \cite{holstein2020conceptual}.

\begin{figure}
    \centering
    \includegraphics[width=1\linewidth]{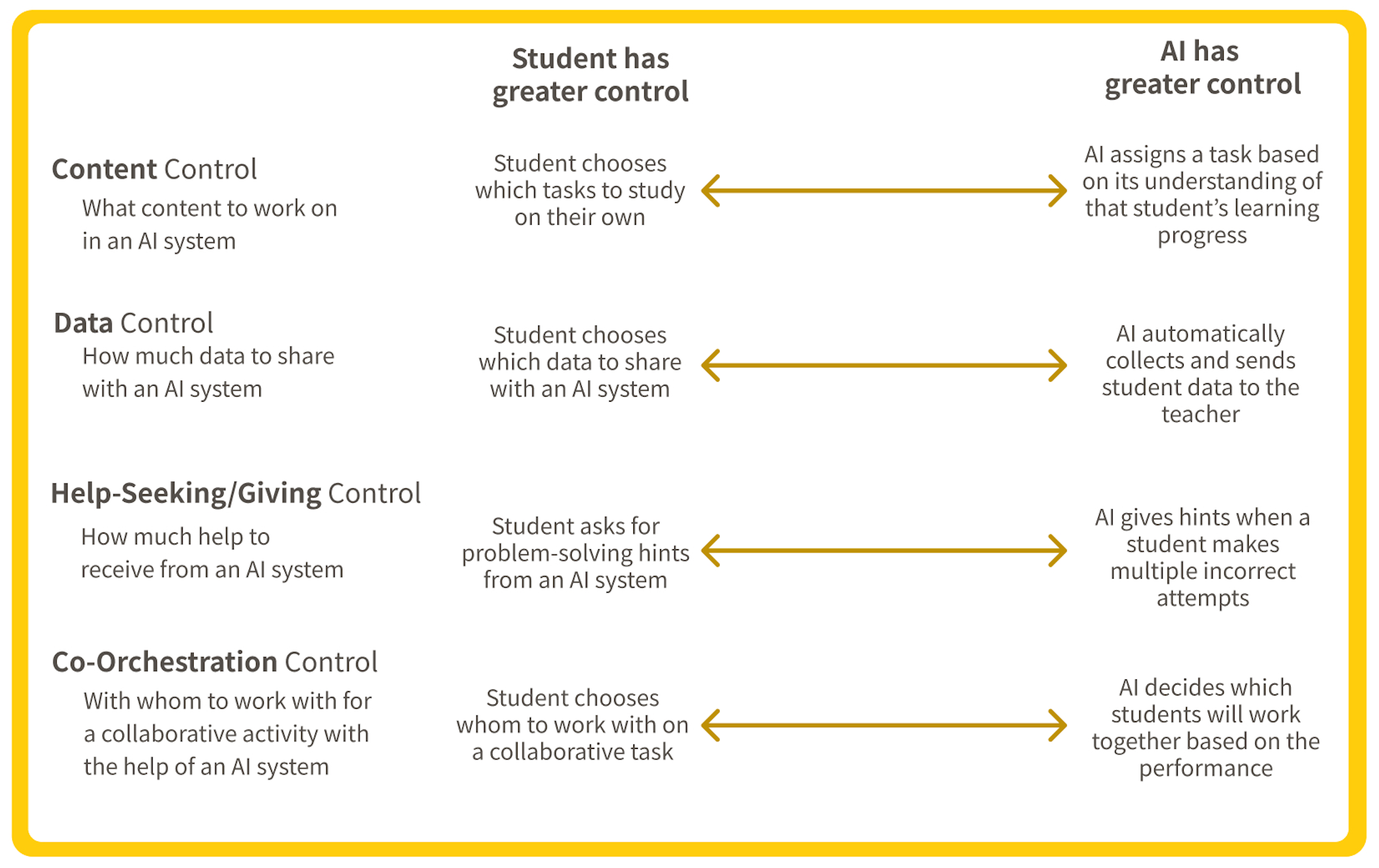}
    \caption{Multi-faceted decision-making control between students and AI systems in the classroom, adopted from \cite{vincoli2025multidimensional} with example cases where one side takes greater control.}
    \Description[Four dimensions of decision-making control between students and AI systems in the classroom]{There are four dimensions of control, namely, Content Control, Data Control, Help-seeking/giving Control, and Co-Orchestration Control. For all four dimensions, we can think about cases where students have more control vs AI has more control. For instance, for Content Control, when students have greater control, they can choose which tasks to study on their own. When AI has greater control, AI would assign a task based on its understanding of the student. For Data Control, student having greater control could choose which data to share with AI, while AI might automatically collect data if AI is given greater control. For Help-seeking control, Students spontaneously ask for help if they have more control, whereas AI might automatically give scaffolding and feedback if AI has more control. Finally, for Co-orchestration control, students with greater control could choose  with whom to work with in a collaborative task, whereas AI could assign pairs if AI has more control.}
    \label{fig:controldimensions}
\end{figure}

How much control to have when learning with digital technology has been empirically investigated before the explicit use of AI \cite{sawyer2017more, snow2015does, nguyen2018student}. Most of these studies on decision-making control with learning technology (without AI) typically considered how much control students vs. technology would have in selecting/receiving tasks in the system (e.g., by testing conditions that differ in the degree of decision-making control), assuming a one-dimensional paradigm \cite{sawyer2017more}. Given the growing prevalence of AI-based technologies, however, the decision-making space today is more complex than a linear progression from low to high control on one dimension \cite{brod2026agency}. Vincoli et al. \cite{vincoli2025multidimensional} proposed that, in today’s AI-supported classrooms, there are four key areas of decision-making agency/control (Figure \ref{fig:controldimensions}): In addition to the traditional dimension of decision-making on selecting content (\textit{Content Control}) \cite{long2014gamification, taub2020agency, long2013supporting, shabana2022curriculumtutor}, studies have looked at \textit{Help-Seeking/Giving Control}, which concerns the balance between the instructional support (e.g., hints and feedback) students seek and the help that the system automatically offers \cite{koedinger2007exploring, nagashima2023promoting, aleven2016help}.  Another line of work examines the degree of control between students and AI on how much/what data to share with the system (\textit{Data Control}) \cite{holmes2022ethics, 10.1145/3772318.3790360,holstein2019designing, khalil2024adaptive, pardo2014ethical}. Finally, Vincoli et al. \cite{vincoli2025multidimensional} highlighted that an increasing number of recent studies have identified the importance of \textit{Co-Orchestration Control}, which refers to the degree of control between students and systems on how much and with whom students want to collaborate during learning \cite{echeverria2023designing, lawrence2022co, nazaretsky2022empowering, olsen2015toward, olsen2021designing, yang2023pair, yang2021surveying}. As AI technology can exercise control over these multiple dimensions of interactions in the classroom, it is essential to understand stakeholder views on such varied aspects of learning and teaching with AI, not just on a specific aspect of decision-making. 

\subsection{Understanding (Mis-)Alignments between Teachers and Students on AI Use in K-12 Classrooms}

In a complex, dynamic environment where multiple stakeholders interact with each other with technology, it is critical to recognize how each stakeholder group perceives the technology, what they expect the technology to do and what they think other stakeholders in the environment would/should do \cite{nagashima2025understanding}. In the classroom, it is possible that what teachers want/expect AI tools to perform is different from what students want/expect AI tools to do. A one-sided view only from teachers, explored in prior work, may hinder effective, learner-centered design of future AI-based learning systems for classroom use. To design effective AI-based tools, it is vital to understand multiple stakeholders’ (different) views, especially from school students’, so that the tool will achieve collective goals in a multi-stakeholder environment \cite{vincoli2026i}. In the paragraphs below, we outline why studying K-12 teachers and students’ perspectives would offer valuable insights.

\subsubsection{K-12 classrooms provide different context than higher education} To date, relatively few studies have explored K-12 students’ views on AI use specifically within the school classroom contexts \cite{chen2020artificial, vincoli2026i}. Although the CSCW and HCI communities are well-positioned to investigate multi-stakeholder perspectives in socially situated environments such as classrooms, prior research has largely focused on teachers, and more recently, university students as primary design partners or participants \cite{10.1145/3711104, 10.1145/3687038, 10.1145/3637358}. These recent studies, some of which surveys both instructors’ and students’ views on Generative AI, offer deep insights into how university students form trust in AI \cite{adnin2025examining, amoozadeh2024trust,walker2024they}, trust in instructors \cite{luo2025does}, and how they metacognitively make decisions about AI use while considering other stakeholders’ perspective \cite{adnin2025examining}. 

We posit, however, that findings from higher education contexts may not directly translate to K-12 classrooms due to several important contextual differences. First, K-12 students may be less able to metacognitively make effective decisions about how to act in ways that support productive interactions with both technology and teachers. Studies in cognitive and developmental psychology have shown that general metacognitive skills develop across the lifespan, with clear gaps between adolescents and young adults \cite{vukman2005developmental}. Moreover, in a multi-stakeholder environment, one needs to take the perspective of others to inform their own decisions  \cite{de2023perspective, nagashima2025understanding, yadollahi2022children}, as demonstrated in studies with university students \cite{10.1145/3711104}. This capacity, however, is also known to mature into adulthood. Prior work has shown that school-aged students are less effective than young and older adults at inferring others’ mental states (i.e., Theory of Mind) \cite{de2023perspective}. 

Second, classroom dynamics in K-12 contexts differ significantly from those in higher education. For example, Kovač et al. \cite{kovavc2010trusting} describe how trust formation between educators and learners in K-12 contexts differ from that in universities. While university students often base trust in instructors on disciplinary expertise and authority, school students develop “personal trust” in teachers, grounded in ongoing, relational interactions that occur every day and perceptions of care \cite{kovavc2010trusting}. These differences may, in turn, influence how students and teachers perceive and negotiate AI use in the classroom. For instance, the relationship between instructors and students in higher education may be more open to challenge by students as it is primarily rooted in disciplinary knowledge, not based on care and kindness by teachers. Additionally, instructors in higher education may hold different expectations of students than those in K-12 settings (e.g., school students lack AI literacy and more vulnerable to biased content and mis-information \cite{han2024teachers}), possibly open to letting university students use AI tools more independently (with responsibilities). Thus, we believe that these contextual factors would affect both stakeholders perceive decision making with AI in the classroom.

\subsubsection{\textit{A need for explicit, fine-grained comparison of stakeholder views in AI-supported K-12 classrooms}.} To the best of our knowledge, there are two studies that focus on both teachers and K-12 students on AI use in the classroom in their design-based research. In one study, teachers and students were introduced to a variety of hypothetical cases of AI use in the classroom and were probed how they would feel in those situations \cite{holstein2019designing}. Another study tested teacher-led, student-led, and system-led pair-matching policies for collaborative work within intelligent tutoring software in actual classrooms and interviewed teachers and students afterwards \cite{echeverria2023designing}. These studies found that, while there were views that were aligned between teachers and students, there were some misaligned views. For instance, they found that students preferred to own greater control over selecting who to work with in a group work. However, teachers often wanted to exercise greater control over grouping decisions and wanted to get alerts on students’ status because they recognized that students cannot necessarily make decisions that are effective for their learning (e.g., friends engage in chats unrelated to learning) \cite{holstein2019designing, echeverria2023designing}.

These prior works show a few focused aspects of teacher-student (mis)alignments, yet fail to cover other key aspects of decision making systematically. First, these studies have a strong focus on Co-Orchestration Control and Data Control, which limit insights into other critical decision-making dimensions in AI-supported classrooms (Figure \ref{fig:controldimensions}) \cite{vincoli2025multidimensional}. Second, they do not compare and contrast views of stakeholders in a fine-grained, explicit manner. The comparison between teachers’ views and students’ views was performed at the very high level in terms of data granularity (e.g., only comparing the most/least preferred topics from each participant group); it is possible that, by explicitly comparing stakeholder views at a lower granularity level in the data (comparing on every single topic emerged from the data in the early phase of qualitative data analysis), some nuanced differences between the stakeholders might emerge (that might be lost if the comparison only occurs at the high level). Further, these past studies do not deeply engage with the sources of misalignments, offering limited actionable insights to overcome the misalignments. 

Our work takes a systematic and holistic approach in covering all the key dimensions of AI use in the classroom to deeply understand views of K-12 students and teachers. As well, we conducted a \textit{pair-matching} data analysis, identifying nuanced differences between their views by explicitly comparing and contrasting across 438 individual themes from speed-dating sessions with teachers and students. Our results demonstrate the dynamic complexity of human-AI interactions in the classroom, where human-human relationships (between teachers and students) have a significant impact on how they perceive and would like to interact with AI systems. We also discuss where the misalignments come from and ideas for filling the perspective gap.

\section{Method}
We conducted a qualitative speed-dating study \cite{holstein2019designing, yang2021surveying, zimmerman2017speed} in Germany exploring students’ and teachers’ preferences on AI use in the classroom using storyboards, with a particular focus on how much control AI or student should take in making decisions during learning (and how much teachers would intervene). Speed dating is a design method developed by Zimmerman and Forlizzi \cite{zimmerman2017speed} in which users are presented with multiple possible futuristic design scenarios. By observing/engaging with how users reflect on each scenario, designers or researchers intend to probe their deep needs and wants around that topic. In our work, we adopted the speed dating method to explore our research question: \textit{What preferences do school students and teachers in Germany have regarding student versus AI decision-making control in AI-based learning systems in the classroom, and how do these preferences (mis)align?} 

\subsection{Participants}
A total of 16 school-aged students (mean age: 14.19, SD: 2.04) and 15 school teachers (mean year of teaching: 8.21, SD: 8.28) individually participated in the study in Germany (Table \ref{tab:participants}). Participants were recruited through direct emails/visits to the school. For both groups, participants chose to participate in the study in German or English. Participants were given a consent form before the study, and only after signing the form (or getting a signature from their parent or legal guardian), they participated in the study. Students received 12 Euros after the study and teachers received 40 Euros. Participation in the study was fully voluntary. The recruited teachers and students had no known connections with each other, and they did not interact in any way during the study. We chose to conduct our study in Germany because our research team has established collaborations with German institutions and local schools, which facilitates efficient access to relevant data and expertise.
\begin{table}
\centering
\caption{Participants' demographics, in the order of participation. "S" and "T" denote student and teacher participant, respectively.}
\label{tab:participants}
\renewcommand{\arraystretch}{1.2}
\begin{tabular}{cccccc}
\toprule
         ID&  \makecell{Age/years \\of teaching}&  \makecell{Grade level\\ (taught)}& \makecell{Subjects taught}& \makecell{Language \\used}&\makecell{Participation \\mode}\\
          \midrule
         S1&  16&  11th& & English&In-person\\
         S2&  16&  10th& & English&Online\\
         S3&  12&  6th& & English&Online\\
         S4&  14&  8th& & German&Online\\
         S5&  12&  6th& & English&In-person\\
         S6&  14&  7th& & German&Online\\
         S7&  13&  7th& & German&Online\\
         S8&  17&  11th& & German&Online\\
         S9&  12&  6th& & English&Online\\
         S10&  16&  10th& & English&Online\\
         S11&  13&  8th& & English&Online\\
         S12&  12&  7th& & English&Online\\
         S13&  16&  10th& & English&Online\\
         S14&  18&  12th& & English&Online\\
         S15&  12&  6th& & English&Online\\
         S16&  14&  8th& & English&Online\\
         \midrule
         T1&  10&  5-12th& \makecell{German, French}& German&Online\\
         T2&  3.5&  1-4th& \makecell{German, Math, Science, \\Music, Sport, French, Art}& German&In-person\\
         T3&  4&  5-12th& \makecell{German, History}& German&Online\\
         T4&  15&  5-12th& \makecell{German, History, \\Dramatic Play}& German&Online\\
 T5& 15& 5-12th& \makecell{German, English, Sport}& German&Online\\
 T6& 7& 8-10th, 12th& \makecell{Math, Physics}& German&Online\\
 T7& 1.5& \makecell{7-8th, \\10th, 12th}& \makecell{Math, CS}& German&In-person\\
 T8& 2& \makecell{5th, 7th, \\10-11th}& \makecell{Math, Physics}& German&Online\\
 T9& 2.5& 3rd& \makecell{German, Math, Music, \\Art, Sport, Religion}& German&Online\\
 T10& 2& 5th, 7-9th& \makecell{Music, Math}& German&In-person\\
 T11& 1& 6-12th& \makecell{Physics, Sport}& German&Online\\
 T12& 0.5& 6-7th& \makecell{CS, Geography}& German&Online\\
 T13& 21& 5-9th& \makecell{English, Social Sciences, \\Communications and\\ Media, Business}& German&Online\\
 T14& 10& 11-13th& \makecell{German, History}& German&Online\\
 T15& 28& 5-10th& \makecell{German, Art}& German&In-person\\ 
 \bottomrule
    \end{tabular}

\end{table}

\subsection{Materials}
\begin{figure}
    \centering
    \includegraphics[width=0.75\linewidth]{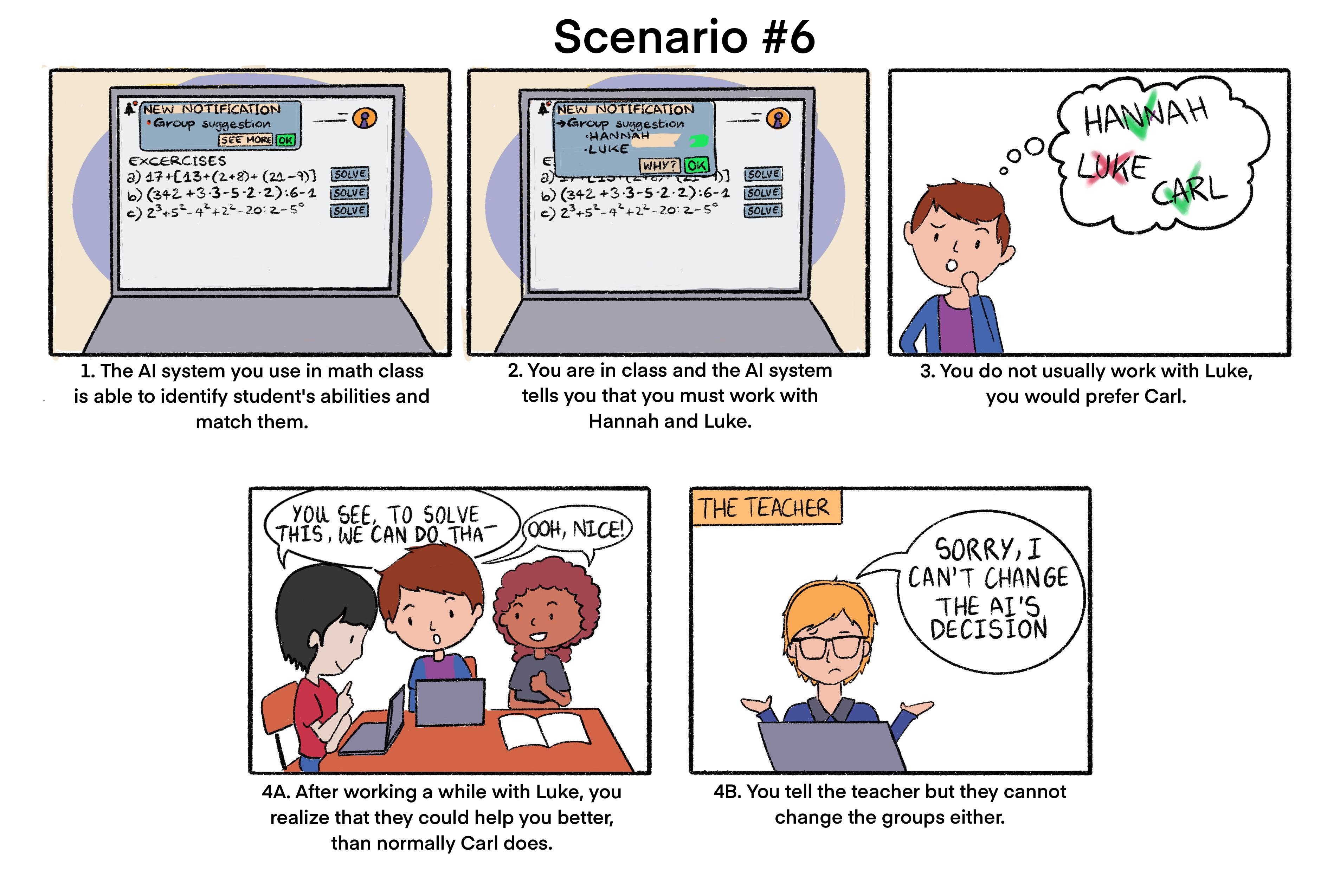}
    \caption{A scenario (storyboard) for students that illustrates the situation where AI has greater control on making decisions to form groups in the class. 4A shows a positive ending where a student (“you”) realized that you can actually work effectively with Luke and Hannah despite your earlier expectation, whereas 4B shows a rather negative consequence where your teacher cannot change the grouping policy, failing to accommodate your (student) request.}
    \Description[A storyboard for students that illustrates the situation where AI has greater control on making decisions to form groups in the class.]{In the first scene, it says "The AI system you use in a math class is able to identify students' abilities and match among students." In the second scene, it says "You are in class and the AI system tells you that you must work with Hannah and Luke." In the third scene, "You do not usually work with Luke, you would prefer Carl.". Then there are two possible endings provided, where one says "After working a while with Luke, you realize that they could help you better than normally Carl does." In another ending, it says "You tell the teacher but they cannot change the groups." }
    \label{fig:studentscenario6}
\end{figure}

To facilitate the speed-dating sessions, we prepared eight storyboards that show potential situations of AI use in the classroom for each group of participants (i.e., eight storyboards for students and another set of eight storyboards for teachers). As shown in Figure \ref{fig:studentscenario6} (one scenario we showed to students), each storyboard has five scenes in total, where the first three introduce the context of the story and actions taken by stakeholders in the classroom (e.g., AI automatically assigns groups, Figure \ref{fig:studentscenario6}). The last two scenes show the ending/outcome of the action, where one (4A) shows a positive consequence while the other (4B) shows a negative consequence. For instance, in Figure \ref{fig:studentscenario6}, 4A illustrates a positive consequence of AI-based automatic group assignment in the classroom where students are working together effectively while 4B shows a negative consequence that the teacher cannot change the grouping despite a student’s request to change it.

\begin{figure}
    \centering
    \includegraphics[width=0.75\linewidth]{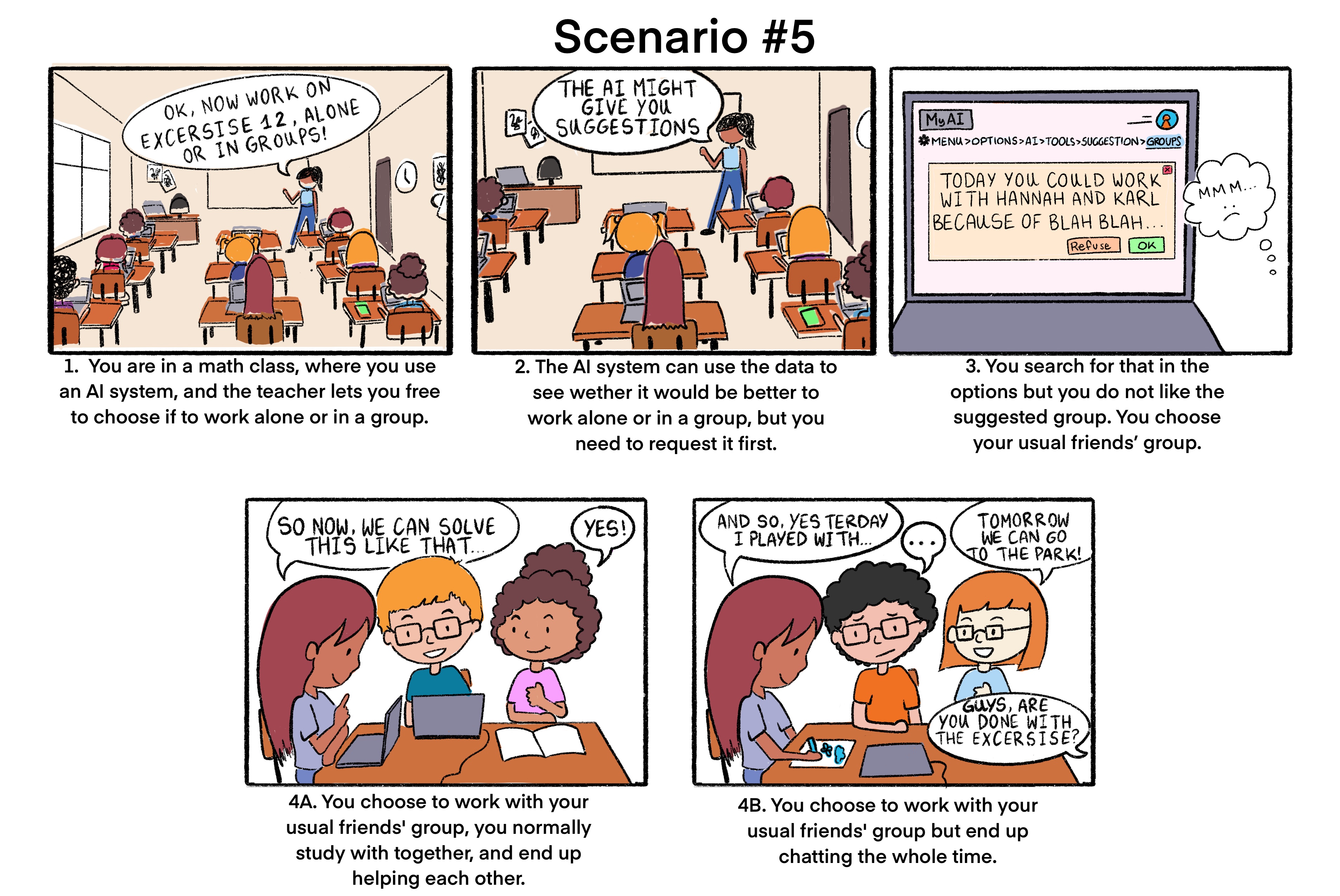}
    \includegraphics[width=0.75\linewidth]{6Scenario_Students.jpg}
    \caption{These two scenarios (for student participants) on Co-Orchestration Control illustrate what might happen when students have greater control over grouping (Scenario \#5) vs. AI has greater control (Scenario \#6). Scenario \#5 shows that students can choose to work with particular students. 4A shows a positive ending where the student-driven grouping led to effective learning. 4B shows a situation where they ended up chatting during the group work without making progress on the task. Scenario 6 is identical to Figure \ref{fig:studentscenario6}.}
    \Description[Two storyboards are presented, which show scenarios on the same Co-orchestration control but illustrate situations when students have more control vs AI has more control.]{For the one with students having more control, in the first scene, it says "You are in a math class where you use an AI system, and the teacher lets you freely choose who to work with in a group." In the second scene, it says "The AI system can use the dat to see whether it would be better to work alone or in a group, but you need to request it first." In the third scene, "You search for that in the system but you do not like the suggested group. You choose your usual friends' group." Then there are two possible endings provided, where one says "You choose to work with your usual friends' group that you normally study with together, and end up helping each other." In the other ending, it says "You choose to work with your usual friends' group but end up chatting the whole time." The one presented that illustrates a situation with greater AI control is the same as the figure presented before.}
    \label{fig:studentscenario56}
\end{figure}

These situations depicted in the scenarios were chosen from similar studies (e.g., studies with teachers using storyboards on AI use in the classroom) \cite{holstein2019designing, yang2021surveying} and based on empirical studies that tested different degrees of student autonomy/AI’s decision making (e.g., a study comparing conditions where half of students can choose what to learn within AI-based software while the other half gets automatically assigned content by AI) \cite{long2014gamification}. These scenarios present features of AI-based systems that have been used in research or practice (i.e., realistic future scenarios) \cite{echeverria2023designing, nguyen2018student}. We structured the scenarios so that each scenario corresponds to each of the eight example cases in Figure \ref{fig:controldimensions}, based on the four dimensions of decision-making control between students and AI in the classroom \cite{vincoli2025multidimensional}. That is, for each of the four dimensions, we created two different scenarios (hence eight scenarios in total) where one gives students greater autonomy in making decisions (\textit{full learner control}) whereas the other gives AI greater control in making decisions (\textit{full AI control}). An example of these two cases (for Co-Orchestration Control) is presented in Figure \ref{fig:studentscenario56}. We chose to use scenarios as a probing strategy because we had known (from prior interactions as a lab) that most teachers and students in Germany had not used AI-based learning systems in the classroom; by offering potential use cases in the form of storyboards, we hoped to elicit insights that would otherwise be challenging to get. Also, unlike past studies where storyboards were not created systematically, we decided to provide scenarios that illustrate two \textit{extremes }(i.e., \textit{full learner/AI control}) and positive/negative endings in order to elicit insights that are as fair as possible (i.e., we wanted to avoid receiving biased responses by showing only e.g., positive cases, or situations where AI has always greater control). In particular, through our prior research experience, we expected that it would be difficult to get deep insights from school students by only presenting \textit{neutral }scenes that might not evoke any strong feelings. 

\begin{figure}
    \centering
    \includegraphics[width=0.75\linewidth]{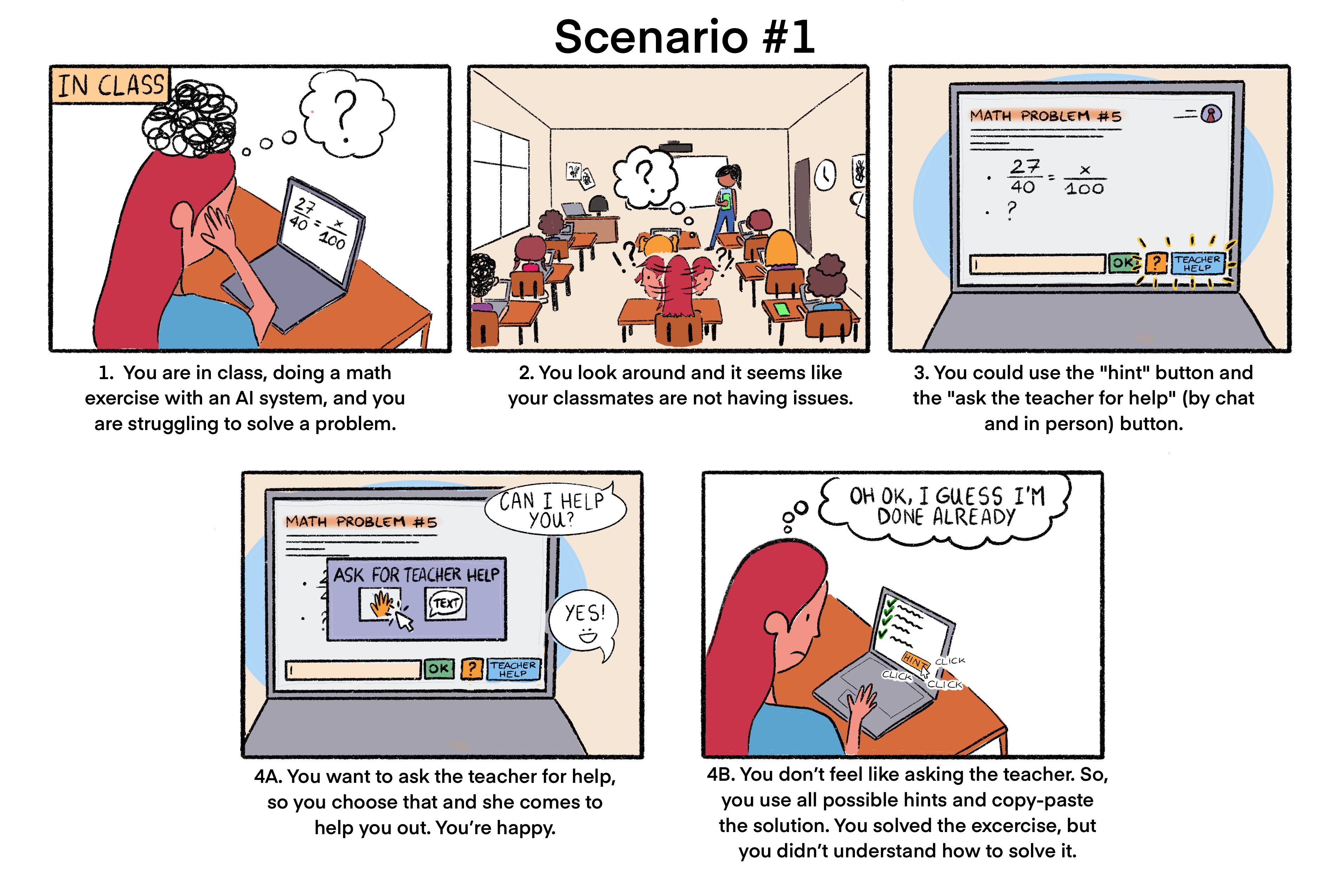}
    \includegraphics[width=0.75\linewidth]{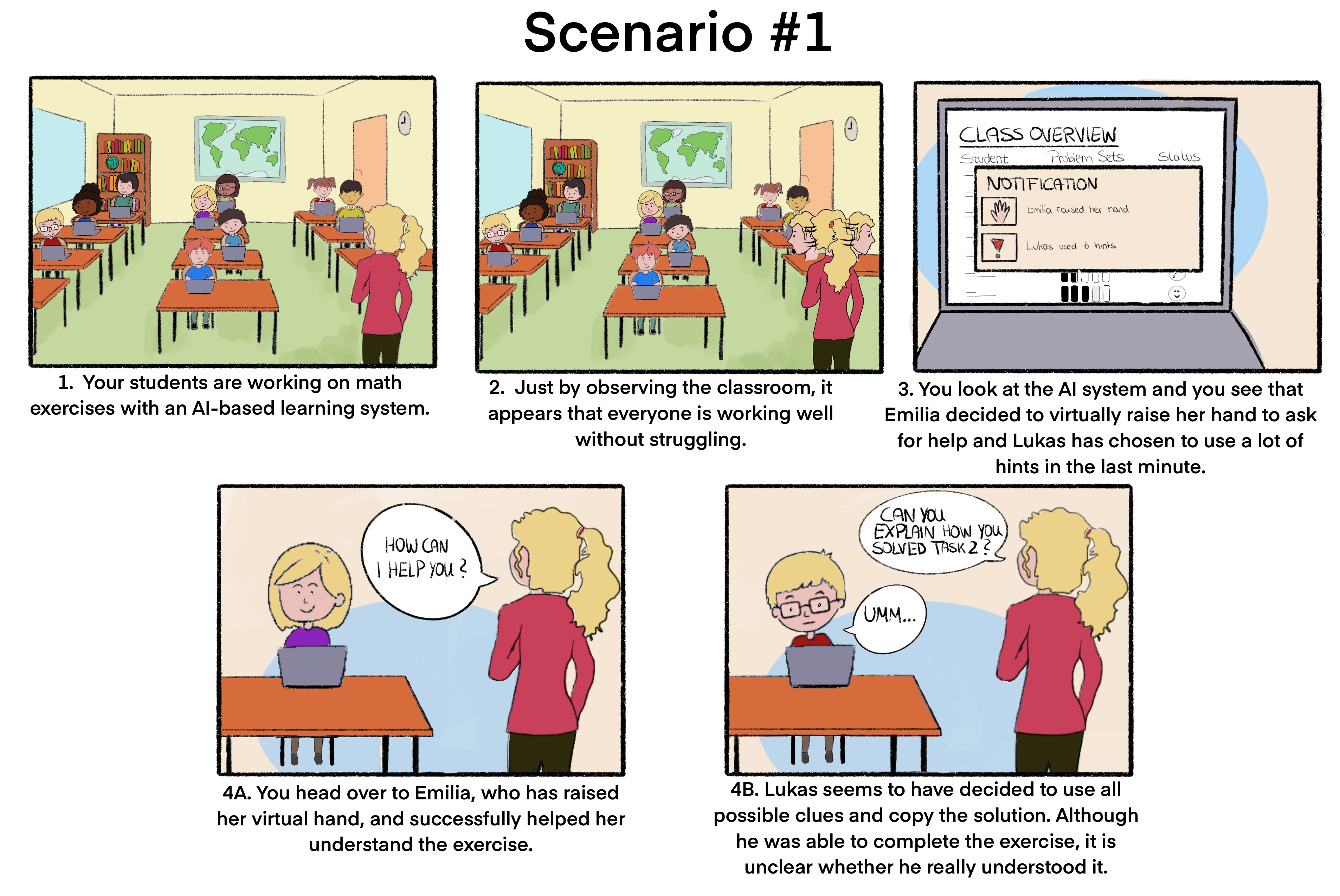}
    \caption{An example of the same scenario, prepared for student sessions and teacher sessions. One designed for students (top) shows the situation where a student (“you”) makes decisions on asking for help (Help-Seeking/Giving Control, \textit{full learner control}) and the one designed for teachers (bottom) shows the same situation from the teacher’s viewpoint.}
    \Description[Two storyboards are presented, which show the same scenario from teachers' and students' perspectives.]{For the one from a student's perspective it first says "You are in a class, doing a math exercise with an AI system, and you are struggling to solve a problem." Second, "You look around and it seems like your classmates are not having issues." Third, "You could use the hint button and the 'ask the teacher for help' button." One of the two endings is "You want to ask the teacher for help, so you choose that and she comes to you to help you out. You are happy." The other ending says, "You don't feel like asking the teacehr. So you use all possible hints and copy-paste the solution. You solved the task but you did not understand how to do so." In the teacher version, it starts with "Your students are working on math exercises with an AI-based learning system." Then it goes, "Just by observing the class, it appears that everyone is working well without struggles." The third scene says, "You look at the AI system and you see that Emilia decided to virtually raise her hand to ask for help, and Lukas has chosen to use a lot of hints in the last minute." One ending says, "You head over to Emilia who has raised her virtual hand, and successfully helped her understand the exercise." The other ending says, "Lukas seems to have decided to use all possible clues and copy the solution. Although he was able to complete the exercise, it is unclear whether he really understood it."}
    \label{fig:teacherstudent}
\end{figure}

The eight scenarios presented to students were nearly identical to those presented to teachers. We first drew scenarios for students from a student’s perspective, then adapted them into a version for teachers, reflecting a teacher’s viewpoint. For instance, as shown in Figure \ref{fig:teacherstudent}. these two scenarios depict the identical situation of students receiving desired vs unexpected help from the teacher but differ only in terms of where the point of view is: one (top) from a student’s point of view and the other (bottom) from the teacher’s point of view. All scenarios were prepared in English and German and are available on OSF (see: \url{https://osf.io/zxh69/}\cite{nagashima2026supplementary})

\subsection{Procedure}
Participants individually attended the session either online (via Microsoft Teams/Google Meet) or in-person in a university lab. Two researchers (with at least one German native speaker) conducted each session. After initial several \textit{warm-up} questions on participants’ age/years of teaching and how they use technology in the classroom for teaching and learning, we introduced the eight scenarios (teachers and students were presented with teacher scenarios and student scenarios, respectively), following the speed-dating approach \cite{zimmerman2017speed}. For each participant, scenarios were presented in a random order to mitigate any potential influence of what participants see earlier in the session on their input for later scenarios. First, participants were introduced to the first three scenes of a scenario (before we reveal two potential consequences) and were asked to imagine and share (verbally) what potential consequences might occur next. We did so to ensure a clear understanding of the scenario context and not to bias participants’ thinking by showing researcher-made outcomes. We then engaged with participant-generated consequences by asking some questions to get deeper insights. After that, we revealed our positive and negative consequences in the scenario and asked participants to reflect on how they would feel if they had been a student or teacher (depending on the participant) in the depicted situations for both outcomes. To scaffold participants’ thinking process on the decision-making control that we were focused on, we also gave some pre-defined guiding questions when we felt scaffolding would be necessary (especially for students). These questions are included in Table \ref{tab:control_dimensions}. We video-recorded all the sessions for later analysis except for one student where the participant agreed to participate but refused to be recorded (we only took notes). Each session lasted about 45 to 60 minutes.

\begin{table}
    \centering
\caption{Guiding questions asked during speed dating.}
\label{tab:control_dimensions}
\renewcommand{\arraystretch}{1.3}
    \begin{tabular}{m{2.4cm}m{5cm}m{5.2cm}}
    \toprule
         \makecell[l]{Dimension of \\Student vs. AI \\Control}& Question for students& Question for teachers\\
    \midrule
         Content Control& How do you feel that this AI-based adaptive system can make decisions about what you should learn?& How do you feel that this AI-based adaptive system can make decisions about what your students should learn?\\
         Data Control& How does it make you feel that your teacher can see your learning data?& How does it make you feel that you can see the student's data?\\
         \makecell[l]{Help-Seeking/\\Giving Control}& How would your learning look different if AI could give you feedback?& How would the student learning look different if AI could give your students feedback?\\
         \makecell[l]{Co-Orchestration\\Control}& Do you think it's important to have the option to work alone or in a group? How would you feel if AI decides for you?& Do you think it's important to have the option to work alone or in a group for your teaching? How would you feel if AI decides for you and for students how to form groups?\\
    \bottomrule
    \end{tabular}
\end{table}

\subsection{Analysis}
Sessions with students generated approximately 14.0 hours of recording, while sessions with teachers produced 17.4 hours. Three researchers (including two German native speakers) first transcribed and coded the data following the open-coding procedure \cite{corbin2014basics}, resulting in 644 and 637 codes for student data and teacher data, respectively. During the initial coding, the entire project team met weekly to discuss the coding procedure in both datasets and resolve any disagreements. Warm-up questions about background information were excluded from the analysis as they were intended only to create a comfortable atmosphere for participants (especially for students). Below, we describe the two major phases of our analysis process (Figure \ref{fig:analysis}).

\subsubsection{\textit{\textit{Phase 1: Separate Analysis of Student and Teacher Data}}.} Within each dataset (students and teachers), five researchers (the three coders plus two data collectors) conducted Affinity Diagramming to find common themes (i.e., we initially analyzed student data and teacher data separately, Figure \ref{fig:analysis}, Phase 1). Affinity Diagramming takes a bottom-up analytical approach that groups codes into hierarchies of themes, making it possible to move from individual codes towards over-arching patterns \cite{holtzblatt2022consolidation}.

We began by grouping codes \textit{within} each of the four decision-making control dimensions (see Figure \ref{fig:controldimensions}), and then \textit{across} control dimensions. For example, in the teacher dataset collected when we used storyboards for the \textit{Data Control} dimension, we grouped 10 codes from 7 different teachers into the low-level theme \textit{“I would like the support of an AI system to help students individually.”} This low-level theme was later grouped with four others (from other control dimensions) into a mid-level theme, \textit{“I think AI can help me individualize help and feedback to students.” }As a result, Phase 1 produced separate sets of low-, mid-, and high-level themes for students and teachers.

\subsubsection{\textit{\textit{Phase 2: Pair-Matching Student and Teacher Themes.}}}
\begin{figure}
    \centering
    \includegraphics[width=0.75\linewidth]{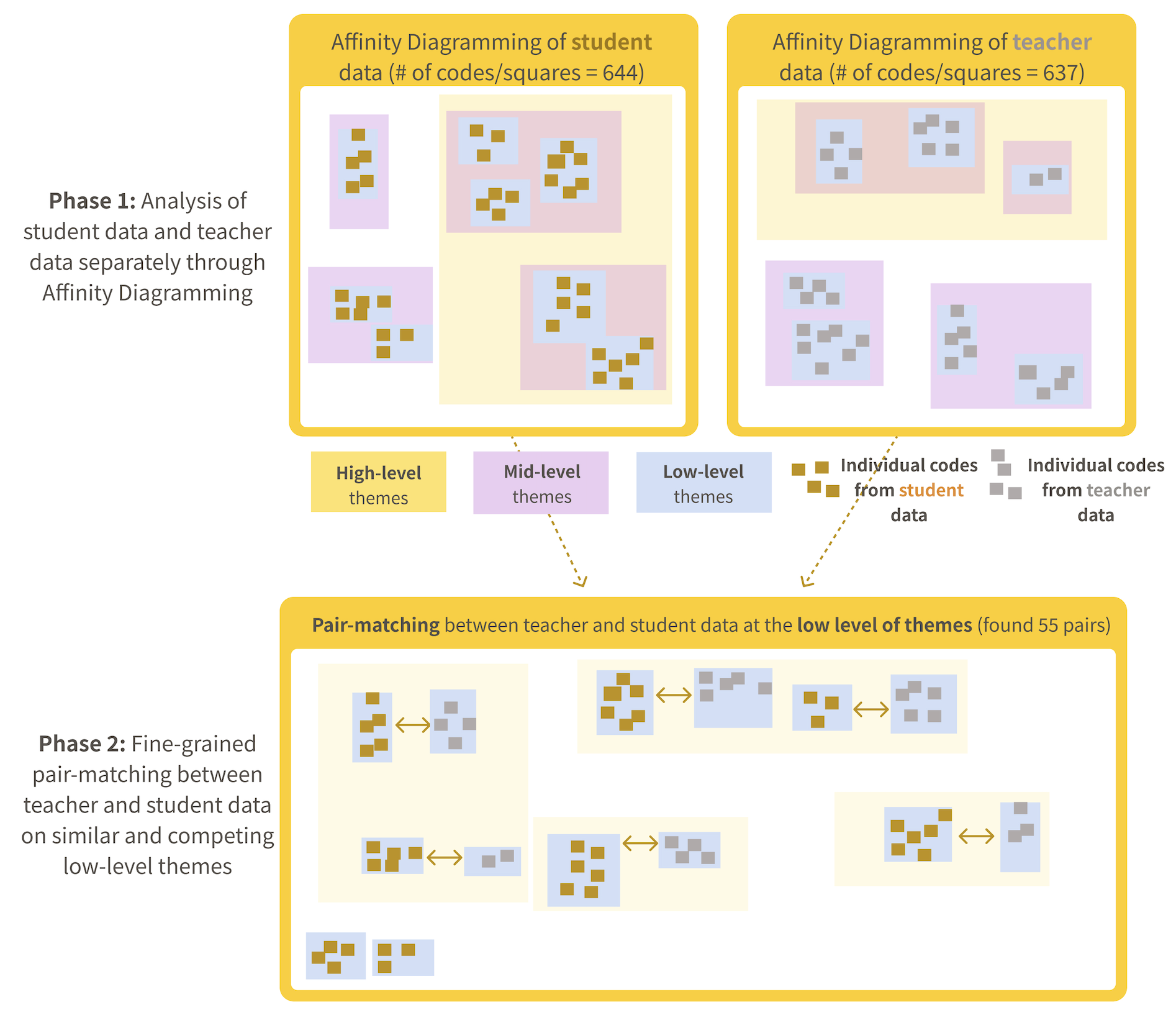}
    \caption{A visual diagram of our analysis process. We first conducted Affinity Diagramming separately on student data and teacher data (Phase 1, top), then used low-level themes from those data to conduct pair-matching analysis in Phase 2 (bottom).}
    \Description[The figure illustrates the two-phase analysis process]{On the top, it has two big squares representing data groups for teachers and students, and within each it shows conceptual diagram of how affinity diagramming in each group was conducted. It has a text area that says "Phase 1: Analysis of student data and teacher data separately through Affinity Diagramming." On the bottom, it shows "Phase 2: Fine-grained pair-matching between teacher and student data on similar and competing low-level themes.", with a diagram that shows how the combined datasets were then analyzed through the pair-matching approach.}
    \label{fig:analysis}
\end{figure}

To directly compare perspectives, we then conducted a \textbf{\textit{pair-matching}}\textbf{ analysis} between student and teacher datasets (Figure \ref{fig:analysis}, Phase 2). This process involved systematically matching low-level themes from student data with low-level themes from teacher data (438 in total). We chose to work at the low-level theme layer, rather than mid- or high-level themes, because matches at higher levels were too abstract for us to reveal meaningful contrasts.

We conducted the matching as follows: 1) For each student low-level theme, researchers searched for either a similar or a contrasting teacher theme on the same topic. 2) When a match was found, the pair was recorded on a virtual white board. For example, the student low-level theme “\textit{My teachers try to help us}” was paired with the teacher theme “\textit{In the classroom, it is my job to help the students learn}” (similar themes). Conversely, the student theme “\textit{Hints provided by the AI system are not very useful for me}” was paired with the teacher theme “\textit{Hints from the AI system can be a useful way to assist students}” (contrasting themes). 3) If no partner theme was available in the teacher data, the student theme was discarded. This procedure allowed us to find how teachers’ and students’ views are aligned or misaligned on each of the important topics we identified within the datasets, instead of only comparing high-level themes, which did not result in meaningful, in-depth comparisons.

Phase 2 yielded 55 matched pairs. We then grouped these 55 pairs into 11 mid-level themes, and subsequently into five high-level themes that synthesize similarities and differences between teacher and student perspectives. Note that although Phase 1 also produced mid- and high-level themes separately for each dataset, we do not report those in this paper. Still, those mid- and high-level themes for each dataset were useful refining low-level themes before moving to the matching process in Phase 2. This iterative analysis process, which took about seven weeks, involved frequent discussions of validity and reliability. Following conventions of Affinity Diagramming, we did not use inter-rater reliability statistics or predefined coding schemes \cite{10.1145/3359174}.

\subsection{Positionality Statement}
Five researchers contributed to this research project. They come from different cultural backgrounds (Germany, Italy, Japan, China, and the US) and represent multiple disciplines (HCI, Computer Science, and the Learning Sciences). Two members of the research team were born and have school experience in Germany, and actively offered insights into values and norms held in classrooms in Germany during the design and analysis. The five researchers on the project have prior experience working with school-aged children and teachers in Europe, North America, and Asia in their research and other work. They also have research expertise with intelligent tutoring systems, which is the platform of focus for the design of the scenarios. It is of note that their past experiences and viewpoints may have influenced the design of the study, study materials, analytical approach, and interpretation. 

Further, while it is not our intention to conduct cultural comparisons, we note that the German school context may differ from other educational contexts, especially regarding technology use (e.g., U.S. schools, where many prior studies were conducted). For instance, PISA 2022 \cite{pisa2023results} shows that U.S. schools have been better equipped with computers than German schools since 2012. In 2012, there were 0.95 devices per student in the U.S., compared to 0.65 devices per student in Germany (OECD average: 0.68). This gap widened by 2022, when U.S. schools had 1.74 devices per student, whereas German schools had 0.55 devices per student (OECD average: 0.81). Also, teachers’ readiness towards technology use is also different between these countries. The same PISA2022 reported that while US school teachers have been better prepared for technology use in instruction since 2012 (above OECD average), German teachers are catching up only recently in 2022 (reported by school principals) \cite{pisa2023results}. Together with our own observation of German and schools in other countries, we consider that our study context might offer unique perspectives that may have affected how students and teachers express their thoughts in the study.

\section{Results}

The initial coding process produced 644 individual codes for student data, and 637 codes for teacher data. After comparing across 438 individual low-level themes from students’ and teachers’ data, 55 pairs of themes were found (Figure \ref{fig:analysis}, Phase 2). These 55 pairs were grouped into 11 mid-level themes, which produced five high-level themes: \textit{Trust in AI}, \textit{Social and emotional aspects of students in the classroom, AI monitoring and feeling of judgement, Hesitation in data sharing and its pedagogical benefits, and Students’ autonomous decision making and help seeking. }In the sections below, we describe these overarching themes, highlighting alignments and misalignment between teachers and students. Whenever a quote is used, we identify their participant ID, age or years of teaching, and which control dimension the quote was referring to (Figure \ref{fig:controldimensions}).

\subsection{Trust in AI and in Teachers }
First, we found a few key similarities and differences in the perception of AI and how much teachers and students would trust AI’s decision making. This theme came up across all the dimensions of decision-making control. In general, students showed mixed feelings when it comes to how much they trust AI’s decisions. While they recognized AI-based systems as a useful technology that assists learning (e.g., “I like that the AI can tell me what would be good for me because it tells me things I do not know yet,” S8, age: 17, Co-Orchestration Control), they also expressed that AI’s decisions cannot always be trusted. This distrust seemed to come from their understanding that AI is incapable of recognizing the appropriate difficulty of learning tasks for each student. For example, one participant commented on trivial mistakes that AI could pick up when they had nothing to do with student’s knowledge: 

\begin{quote}
“There is like one concern I would have if they correct [my input] wrong, there could always be a small issue in the system or maybe you accidentally typed the wrong thing, which [have] happened to me once you typed the wrong thing and then [the AI system] chooses [the next task] automatically after like 10 seconds” (S16, age: 14, Content Control).
\end{quote}

Students’ reluctance in trusting AI also comes from their conceptualization that AI cannot capture social dynamics (e.g., friendship) and personal feelings that students have: “The AI system can’t see my emotions. So if I feel bad […] and I only want to talk and do things with my usual friend group, that the AI can’t see” (S10, age: 16, Co-Orchestration Control). As they recognized that AI is imperfect, students emphasized the importance of their own decision making, and that of human teachers as a \textit{gatekeeper} of the classroom and that AI-based tools should assist human teaching. Students commented that their teachers have a “better connection” to their students and therefore human teachers would understand what students really need, suggesting that teachers understand students as a whole, not just by the information detected by the systems.

Similar to students, teachers also shared that they would not fully trust AI’s decision making. This concern comes from the perceived risk of incorrect assessment of students made by AI-based systems (i.e., assigning tasks that do not match students’ actual knowledge level). Here, teachers emphasized the importance of being able to intervene and override AI’s automatic assignments: 

\begin{quote}
“There are children for whom a certain topic or task is not so suitable. The child may be frustrated afterwards because they cannot solve the task because it is perhaps more difficult than you would have thought the child would be able to do […], so I think it’s important that you have the opportunity to change something. It’s often the case that you think the child can do it and then you realize ‘oh, it’s harder than I thought’ and then you have to take a step back” (T9, years of teaching: 2.5, Content Control).
\end{quote}

Contrary to students’ thoughts that they would trust human teachers over AI systems, many teachers were worried that students would trust AI more than teachers themselves. One teacher stated that students would likely accept decisions made by a machine than decisions by human teachers: “I think that there is even more acceptance among students when the computer says, ‘Here, work with this because it can help you,’ than when the teacher says something that the teacher has consciously decided” (T7, years of teaching: 1.5, Co-Orchestration Control). Teachers thought in such a way because students might think that teachers make decisions with certain intentions (e.g., when forming student groups): “Perhaps [students] would trust [AI’s decision making] more than the teacher, because the teacher only aims to ‘separate the sources of the fire’ [would not put students who dislike with other in the same group]” (T5, years of teaching: 15, Co-Orchestration Control).

In summary, students and teachers both considered that AI is still imperfect in detecting students’ knowledge and preferences and assigning appropriate tasks, therefore preferring more human control. However, they differ on how much students would trust AI vs. teachers. While students emphasize their trust in human teachers, teachers were concerned that students might find AI’s decisions fairer and more reliable because teachers might make decisions with certain intentions that are not preferred by students.

\subsection{Social Dynamics and Emotional Aspects of Students in the Classroom  }
Findings also revealed that both students and teachers emphasized the importance of social and emotional aspects of learning in the classroom and that they recognized that AI cannot capture such nuanced dynamics. First, students were concerned that AI cannot capture personal preferences (e.g., whom they want to work with) and feelings (e.g., how students feel on a particular day). Students mentioned this incapability of AI in handling social and emotional aspects when interacting with scenarios for the Co-Orchestration Control, where students would be grouped together for a task: 
\begin{quote}
“It’s just a little hard for an AI system to know how well you get along with these people. So this personal connection that I think is important […] is just lacking. I guess it’s just dependent on who you’re paired with and how the AI system sees it” (S14, age: 16, Co-Orchestration Control).
\end{quote}

Students would not care about whether they are \textit{appropriately} matched with someone else based on the performance/knowledge level; they prioritize human-to-human relationships much more, advocating for more human control. This insight is well aligned with prior findings in the literature \cite{echeverria2023designing} that emphasize the importance of social dynamics. 

Similarly, teachers were also worried that AI systems would not be able to detect social dynamics and personal feelings of students in the classroom, confirming prior findings \cite{echeverria2023designing}. Unlike students, however, teachers were not only worried about the possibility of AI creating groups where students do not like each other but also about the possibility of creating a situation where some students do not contribute: “So when I look at our class […], it’s not going to be the case that everyone works well, but there will be a quarter of the students who don’t want to, who try to sabotage the whole thing” (T14, years of teaching: 10, Help-Seeking/Giving Control). Given that, teachers would like to be able to intervene AI’s decision making by adding other factors that the system can base its decisions on (e.g., how well certain groups have worked in the past).

Teachers (but not students) also shared a concern that the use of AI or digital devices in the classroom in general would lead to less social interactions within the class. Teachers had a fear towards overly increased use of digital products and AI-based technology in their instruction (while appreciating their benefits). For instance, when seeing a virtual hand raise in one of the scenarios where students can get teacher’s attention through sending notifications via the system (without being noticed by other students) \cite{holstein2019designing}, a teacher said: “So [raising hand] can happen virtually, I think that’s kind of artificial in a group situation like this, where everyone is glued to their screens […]. So a bit of normal interpersonal communication would be good” (T1, years of teaching: 10, Help-Seeking/Giving Control). This finding is notable as the same concept of “invisible hand raises” has been well-received by school teachers in North America \cite{holstein2019designing}. For some teachers, working with students through AI-based systems would make them feel unnatural: “Do I now work with a robot that I can somehow optimize? The emotion is an important thing that needs to be there for the learning success” (T4, years of teaching: 15, Help-Seeking/Giving Control), emphasizing the need for more conversations, communications, and interpersonal interactions in the classroom with students outside of technology.

To summarize, we found that both students and teachers think that AI cannot capture social dynamics and emotional feelings and want to exercise control, which supports prior findings \cite{echeverria2023designing}. Teachers, unlike students, were also concerned about a possible decrease in interpersonal interactions and communications between teachers and students.

\subsection{AI Monitoring and Feeling of Judgement}
Students and teachers alike mentioned the risks and uncomfortable feelings that AI monitoring would cause in the classroom (across different dimensions of decision-making control but most frequently on Data Control). Specifically, students were concerned about the possibility that their data might be shared with people other than teachers (e.g., parents and peers), which aligns with prior findings \cite{holstein2019designing}.

We also found that students already had a certain level of pressure in the school environment: “You’re told you’re going to school under pressure, you can’t concentrate, and the only thing you’re told by the teacher is that you can’t do this and that well, and that you should do this and that better” (S4, age: 14, Data Control). Such a pressure that students already have in the school environment makes it challenging for them to accept another type of consistent monitoring in the school.

Although some teachers showed a positive attitude towards AI monitoring as it may motivate students, many teachers were concerned about over-monitoring caused by AI systems in the classroom. This concern stemmed not only from its potential negative consequences on student feelings (similar to what students shared) but also its consequence on teachers’ own behaviors in the classroom. For instance, teachers mentioned that having AI that notifies teachers who are struggling in the class would change teachers’ behavior: 
\begin{quote}
“It should not look as if you are coming to the student because you are now placing them under general suspicion [that the particular student is struggling], but even in the other scenarios it should actually be that you move around the room and consult with them from time to time so that this does not lead to such an unpleasant situation for the student” (T3, years of teaching: 4, Help-Seeking/Giving Control).
\end{quote}

This statement shows that teachers care about how students would perceive the teacher and want to ensure that they create a learning environment where no students feel embarrassed by making mistakes. Relatedly, teachers considered that over-reliance on AI would risk teacher-student relationships that they have built in the classroom, and that teachers would be able to intervene to help students feel safe: “You have to take away the fear of making mistakes from the students on an interpersonal level. No machine, no AI can do that, you just have to be there as a motivator” (T4, years of teaching: 15, Data Control). Teachers, especially those with many years of teaching experience also mentioned that whether students have a feeling of fear towards AI monitoring would “also depends on the teacher’s personality or the fundamental relationship of trust” (T15, years of teaching: 28, Data Control), indicating that if students trusted teachers, the AI monitoring would not cause a big issue in the classroom.

In summary, we found that students were generally concerned about the unnecessary pressure created by AI systems through monitoring and judgement in the class. Teachers also shared this concern, but also considered the importance of keeping trust between teachers and students and insisted on the value of creating a learning environment where no one should fear of making mistakes and struggling with learning. Provided that there is a good relationship between students and teachers, teachers believed that AI monitoring could be helpful for instruction.

\subsection{Hesitation in Data Sharing and its Pedagogical Benefits}
Students and teachers shared somewhat varying opinions regarding data sharing through AI. We found that students had a strong desire for keeping their own data private and did not want to share it with others unless it is necessary to do so, in line with the findings above on AI monitoring and prior work \cite{holstein2019designing}. Students, however, were willing to let their own teachers to have access to it as they recognized the benefits of sharing data with teachers (but only with their teachers): “If you leave everything to yourself, then no one can help you […], but if you just share the biggest part with the teacher, then the teacher can also help you” (S7, age: 13, Data Control). In both cases, students preferred to have transparent information of what kinds of data are shared and to whom they are shared, as well as how the data are used.

Teachers insisted on a strong desire for having access to data due to its importance for effective instruction: “To what extent is teaching and learning a democratic process? As a teacher, you also have a certain task to fulfill and that certain things only work if they are based on reciprocity” (T1, years of teaching: 10, Data Control). While teachers recognized that students might not feel comfortable sharing their data, teachers stated that with communication and transparency, students would understand why sharing data is essential: “In the computer room we also have screen monitoring. It would probably be good to tell the children so that they are aware of [getting observed]. I think they will agree with that” (T12, years of teaching: 0.5, Help-Seeking/Giving Control). Also, teachers would not like to let students make decisions on data sharing because such student autonomy would make the system less useful for teachers: 
\begin{quote}
“I strongly assume that the low-performing students in particular don’t want to share their data and that is not an overview for me if only the good ones share it or those who want to share it” (T5, years of teaching: 15, Data Control).
\end{quote}

Some teachers, however, would not necessarily like to be fully transparent. For instance, one teacher shared, “If Lukas [the hypothetical student in the scenario] does not know that the teacher will be notified, he may be able to work without a certain amount of pressure” (T9, years of teaching: 2.5, Help-Seeking/Giving Control), thinking that students might prefer not to be told that they are being monitored.

To summarize, students and teachers expressed different aspects of data sharing. Students prioritized privacy and preferred to control who accesses their data, sharing it only when necessary. Teachers emphasized the importance of having access to student data for effective teaching. They were generally reluctant to give students full control over data sharing, fearing that it would limit the effectiveness of the system. However, some teachers (not all) understood that students might have concerns and believed that clear communication about data sharing could help address their discomfort and understand the importance of data collection/sharing through AI.

\subsection{Students’ Autonomous Decision Making and Help Seeking}
Students and teachers also shared several notable perspectives on students’ autonomous decision-making skills and help-seeking behaviors with regard to the use of AI (related to Help-Seeking/Giving Control and Content Control). We break down these subtopics below.

\subsubsection{\textit{Decision-Making Skills and Preferences}.}In general, students would like to own a certain degree of decision-making control over what they do across different decision-making domains. For instance, many students stated that they would appreciate the freedom of making choices on what to learn because they might want to deviate from AI’s recommendations sometimes, recognizing that AI cannot assess students accurately: “If I have a day where I just don’t want to do anything at all, then I would be happy with the easier tasks” (S1, age: 16, Content Control). However, multiple students also expressed the concern that they would not be able to make good choices if they were given the freedom of decision making, recognizing that they may not be able to make appropriate metacognitive judgment on learning: “I can see a lot of students just picking the easier way just because they want more free time or something” (S15, age: 12, Content Control). They considered such a behavior undesirable and would like to get guidance from human teachers: “Our teacher would say which one we should do […] because she knows maybe which ones are easier and which ones are harder” (S16, age: 14, Content Control).

Teachers believed that some students would be able to make effective choices, and offering decision-making opportunities would motivate students to learn: “This means that nothing is dictated by the teacher or specified by the AI, but the students have a choice. This is always good from a motivational point of view” (T6, years of teaching: 7, Content Control). At the same time, similar to students’ perspective, teachers also thought that some students would not make appropriate decisions:

\begin{quote}
“In the worst case, [students] really only practice what they can do, because then they can do the tasks quickly and somehow collect much more positive feedback without making any learning progress, because if you can already do something and keep practicing, then there’s no learning gain)” (T10, years of teaching: 2, Content Control).
\end{quote}

In summary, we found that students and teachers had similar views towards students’ autonomous decision-making skills. They both believed that, while giving greater autonomy to students in making decisions is beneficial, some students would just try to \textit{get by} without engaging with the most appropriate tasks that will help them learn. They both also shared a need for teacher intervention to guide students’ decision making and when the decisions are not made appropriately, to correct them.

\subsubsection{\textit{Help-Seeking with Teachers and AI}} Overall, we found that students would like to decide by themselves when and how to get help in learning with AI, but they emphasized the central role of the teacher in supporting students. One student shared: “The teacher coming around and helping you could help you understand things a lot better and make you understand where the problem is and so they can explain it better than a hint [in AI-based software] can” (S14, age: 18, Help-Seeking/Giving Control). However, students reported challenges and varied experiences in receiving help in the classroom. Some shared that most teachers do not provide help unless students ask for, aligned with PISA data showing less support is provided in German classrooms, compared to the OECD average \cite{pisa2023results}. Despite the lack of sufficient opportunities for getting human help, students shared that the teacher could help them the best: “I think talking with a human can help you better [than an AI system], because it’s more of a dialogue and they can get if you’re understanding” (S13, age: 16, Help-Seeking/Giving Control). As well, we also found that how much students would like to ask for help through the AI system or human teachers may depend on student’s relationships with their teachers: “I do not like my current teacher which is why I do not really communicate with them” (S10, age: 16, Help-Seeking/Giving Control).

Students saw AI as a promising tool for bridging the discrepancy between what students need and what teachers provide in terms of help-seeking and help-giving. In particular, for students who may be shy or hesitant to ask for help, students thought that AI systems can provide alternative ways for students to ask for help, such as digital hand-raising \cite{holstein2019designing} or messaging with the teacher via a chat system.

Teachers also considered that AI has potential to enhance the efficiency and effectiveness of providing help in the classroom. They valued AI for its ability to provide a comprehensive view of student performance (“So I think it can actually be a fundamental help when I see that the students have difficulties here and there, especially in larger classes where you don’t always have an overview of everything,” T11, years of teaching: 1, Help-Seeking/Giving Control). Teachers also recognized that there are students in class who are more hesitant to ask for help in person, and AI tools could serve useful for such students. According to teachers, students’ willingness to ask for help varies, with some students feeling uncomfortable getting all the attention in the class. This speculation matches what students actually said: e.g., “I think that if I ask [for help] in front of the others they would think I am stupid or not smart enough” (S15, age: 12, Help-Seeking/Giving Control). Relatedly, teachers in general considered that they are (but not AI) responsible for helping students succeed (“As a teacher, you have to stay there and make sure that the higher-achieving pupils don’t dictate everything, so that the lower-achieving child doesn’t fall away and simply say yes and amen to everything”, T2: years of teaching: 3.5, Co-Orchestration Control).

In summary, both students and teachers recognized the importance of feedback and help in the classroom learning process. Their views are well aligned; while AI is seen as a tool that can augment teachers’ efforts and offer support efficiently, human-human interactions between students and teachers are still irreplaceable.

\subsection{Where Alignments and Misalignments Lie between Students and Teachers }
\begin{figure}
    \centering
    \includegraphics[width=1\linewidth]{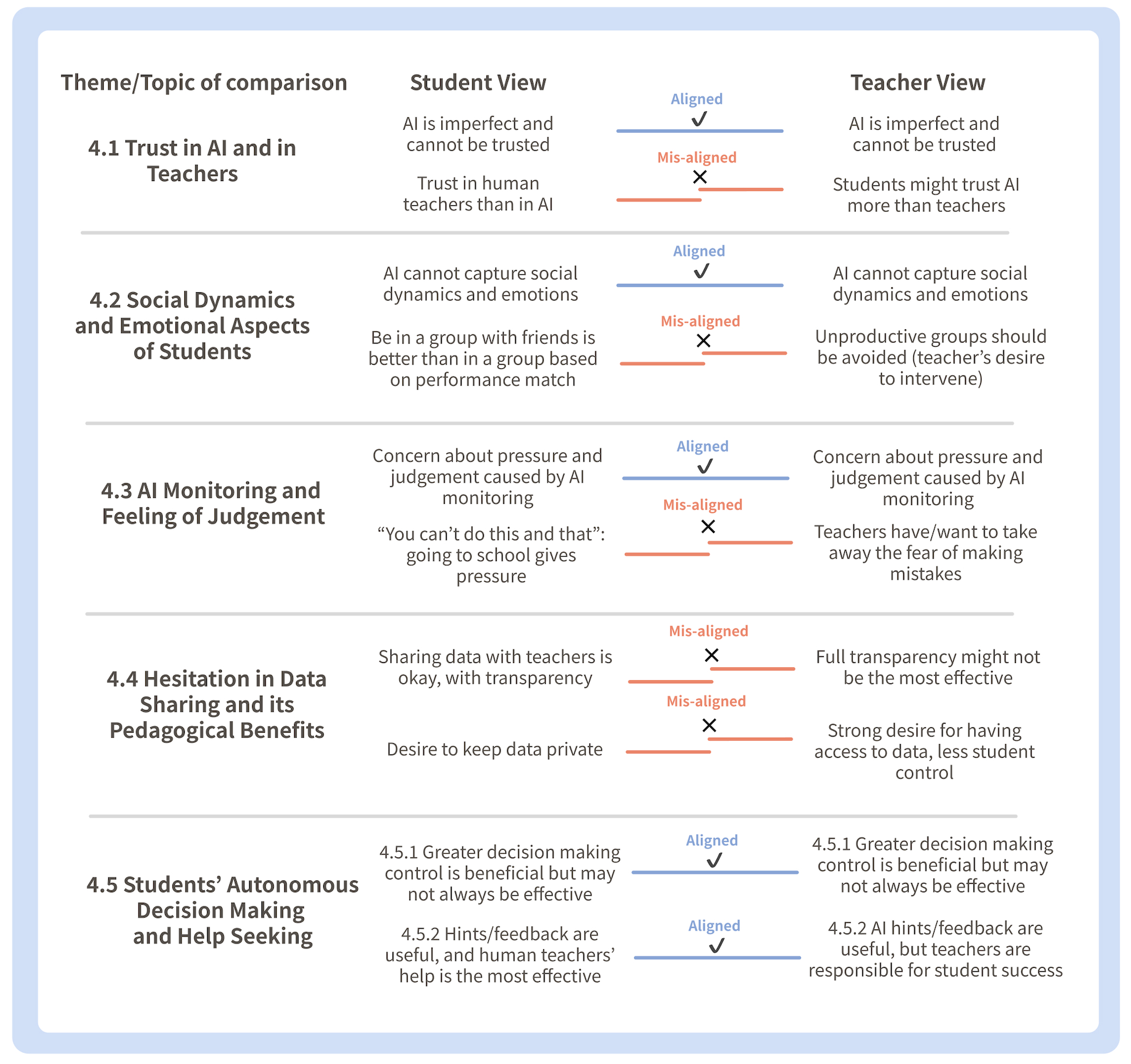}
    \caption{A visual summary that shows what are aligned and misaligned between teachers’ and students’ views.}
    \Description[Summary of key findings]{It shows what key findings the study found, for each presented theme of Trust in AI and in Teachers, Social Dynamics and Emotional Aspects of Students, AI Monitoring and Feeling of Judgement, Hesitation in Data Sharing and its Pedagogical Benefits, and Students' Autonomous Decision Making and Help Seeking. For each, it has "student view" and "teacher view" columns that present what was aligned and misaligned views among the two stakeholders.}
    \label{fig:trustgap}
\end{figure}

The results above illustrate nuanced similarities and differences between students’ and teachers’ views on five major themes generated around AI use in K-12 classrooms. To effectively communicate the main perspectives that are aligned and misaligned, we have developed a visual summary (Figure \ref{fig:trustgap}) that shows how their views are similar/different for each topic of AI use we have observed in the data. As detailed in Figure \ref{fig:trustgap}, students and teachers had some shared preferences and concerns around AI use regarding trust in AI, AI’s detection capabilities, AI monitoring, and students’ autonomous decision-making skills. Yet, it highlights some major misalignments across the topics. Although each of the misalignments is different from each other, one over-arching theme that cuts across the topics is that students tended to prioritize decisions that would make their learning easy, comfortable, and private (e.g., forming groups with friends, avoiding pressure). Teachers, on the other hand, shared opinions that would help achieve effective learning in a collaborative, friendly classroom environment (e.g., creating an environment where students do not have to be afraid of making mistakes). Their perspectives imply that teachers care not only about individuals but also the entire classroom climate. 

\section{Discussion}

\subsection{Key Findings }
As we see an increasing integration of AI tools in classrooms where complex interactions occur, it is essential to deeply understand both students’ and teachers’ perspectives. Through the explicit theme-matching analysis at the low granularity level in the data, our speed-dating study found that while students and teachers shared similar concerns and desires on key topics of AI use (e.g., the perception of imperfectness in AI’s decision-making algorithm and detection capability), their views were not aligned on several notable topics. For example, we found that students tended to trust their teachers more than AI while teachers considered that students might trust AI more due to its “fairer” assessment than humans’ decision making. Within HCI, research has shown university students’ trust toward human explanations (over AI \cite{walker2024they}, but see \cite{luo2025does}), and that they do not over-trust AI \cite{amoozadeh2024trust}. Our study shows a similar pattern among school students, yet contributes a novel misalignment with school teachers’ views. Understanding these misaligned views is key to the effective and sustainable use of AI in the classroom; if teachers hold the misaligned view that students would trust AI more than teachers (when that is not true from a student’s perspective), teachers might less frequently provide scaffolding and active instruction that students would actually hope to get more frequently. 

The findings also suggest that identifying the appropriate degree of decision-making control between the stakeholders is challenging. We found that students would make decisions considering other factors, including how much they trust their own teachers and how much they know about the AI’s capabilities (e.g., in capturing social dynamics). As well, teachers would intervene in this decision-making space by encouraging students to make their own decisions as it would motivate students (c.f., 4.5.1) but they may rather want to provide more scaffolding for students who are not performing well. These insights offer complicated yet meaningful implications for self-regulated learning and AI use in the classroom. More specifically, the findings challenge the theoretical model of “hybrid human-AI regulation” \cite{molenaar2022towards} which implies a gradual withholding of adaptive support (i.e., provide more scaffolding initially, less later in the learning process). Our work suggests that decision making during learning is more complex than a linear progression and requires careful consideration of sociotechnical trade-offs. In particular, it is necessary to evaluate the design space where AI’s technical affordances intersect with the social dynamics among the multiple human stakeholders in the classroom. For instance, while AI may offer objective decisions for supporting students, over-reliance on AI could weaken the essential trust between human stakeholders, which could be fostered not necessarily only through accurate decisions but rather care and empathy (Theme 4.1 \textit{Trust in AI and Teachers}). Similarly, AI’s performance-oriented algorithms, while proven effective in identifying potential matches for grouping students \cite{yang2023pair}, often ignore mood-related learning behaviors and social relationships, which human actors consider (more) important in that decision making scenario (Theme 4.2 \textit{Social Dynamics and Emotional Aspects of Students in the Classroom}). Interestingly, however, while AI monitoring would make students feel surveilled, a positive insight from teachers suggests that trust between teachers and students could mitigate that feeling (and make use of AI monitoring) (Theme 4.3 \textit{AI Monitoring and Feeling of Judgement}). Therefore, the trade-offs could also depend on how human stakeholders form their relationships. 

Overall, the findings illustrate that teachers often care about AI’s potential impact on the entire classroom as a dynamic environment, human-human communications, and the relationship between teachers and students. On the other hand, school students tended to speak for themselves, discussing what they would/would not like. A repeating theme in the results was that the preferred degree of control and trust in student-AI interactions would depend on student-teacher relationships and trust between them. Such a finding underscores the importance of understanding how teachers and students build their relationships when AI is \textit{not} in use, not only when AI is in use. 

These findings both extend and offer insights that differ meaningfully from prior work in CSCW, HCI, and AI in Education. First, consistent with prior research on trust in AI among young adults \cite{adnin2025examining, walker2024they}, our findings show that school students expressed a similar directional preference in trusting humans over AI in classroom contexts. Meaningfully, however, the current study extends prior work conducted within higher education by revealing some more nuanced differences in students’ perspectives. For instance, Han et al. \cite{10.1145/3711104} found that university students would carefully change their behaviors as they recognize how AI would try to collect/interpret data. In contrast, students in our study did not engage in such metacognitive decision making (e.g., by reasoning about others’ perspectives) and rather prioritized their own desires. Furthermore, our results provide deeper insights into the complex decision-making preferences that emerge from the interplay between AI’s technological affordances and contextual factors, including human–human relationships among classroom stakeholders. Prior studies examining both educators’ and learners’ views on AI in K-12 classrooms \cite{echeverria2023designing, holstein2019designing, han2024teachers} and in higher education \cite{adnin2025examining, barrett2023not, luo2025does, wu2024reacting, zastudil2023generative} have not offered similarly detailed accounts of how decision-making preferences are negotiated and expressed. We believe that this contribution is partly enabled by our methodological approach. Specifically, the use of explicit, scenario-based prompts grounded in prior work helped participants envision concrete classroom situations, and our pair-matching analytic strategy allowed us to compare and contrast stakeholder perspectives at a fine-grained level of the data.

\subsection{Where do the Misalignments Come from? }
An important question to reflect on is: \textit{where do the misaligned views come from?} Our findings suggest that the observed misalignments may arise from deep structural and relational factors within the classroom context. Uniquely situated in the K-12 classroom context, we argue that the following potential factors may have influenced the observed misalignments.

\subsubsection{\textit{Developmental and Metacognitive Differences}} First, the differences in the views may partly come from the developmental differences in socio-cognitive skills between school students and teachers/adults. Developmental and cognitive psychology studies have shown the gradual maturation of metacognitive skills as one gets older \cite{zhou2019metacognitive, veenman2004relation, vukman2005developmental}. Specifically, in our study context, the tendency of students speaking for themselves when sharing their thoughts might be attributed to the lack of perspective taking skills, an important socio-cognitive skill that is known to develop during adolescence (13-18 years old) \cite{de2023perspective}. Perspective taking involves understanding other stakeholders’ thinking, viewpoints, and feelings \cite{de2023perspective, nagashima2025understanding}. In our context, we found that teachers often adopt a broader perspective than student participants, trying to take the perspective of students (e.g., having AI that notifies teachers which students are struggling would change teachers’ behavior as they care about what students would guess). Although this tendency might just be due to teachers’ unique pedagogical knowledge of handling a class of students, it is noteworthy that such a difference in perspective taking between educators and learners is seen less often in studies conducted within higher education settings \cite{10.1145/3711104, 10.1145/3637358} (e.g., in their research, university students tend to be able to clearly think about the consequences of using AI for stakeholders other than themselves).

We further posit that some of the misalignment in perspectives may relate to students’ still-developing metacognitive skills in learning-related decision making. Extensive research on self-regulated learning suggests that self-regulated learning involves multiple stages of metacognitive self-assessment, self-monitoring, and self-evaluation \cite{zimmerman2003motivating}. Through these processes, learners make informed decisions about what and how to learn by assessing their abilities and task demands, monitoring their performance, and reflecting on outcomes to adjust strategies as needed \cite{zimmerman2003motivating, nagashima2023promoting}. In our study, students expressed preferences for selecting easier content when given a high degree of decision-making control \cite{long2014gamification} and for forming learning groups based on social compatibility rather than performance level. While these preferences imply their desires about what makes them comfortable during learning, the self-regulated learning literature emphasizes the importance of setting appropriately challenging goals and engaging in iterative cycles of metacognitive assessment, monitoring, and evaluation to support effective learning \cite{long2014gamification, theobald2021self}.

Prior work shows that core self-regulated learning skills (e.g., goal-setting skills) continue to develop throughout adolescence \cite{theobald2021self}. Therefore, it is possible that students in the study generally had not yet developed sufficient level of self-regulated, metacognitive decision-making skills that would support their effective learning; instead, some of their desires reflect “get-by” strategies, which may reflect their wish for bypassing challenging learning situations.

\subsubsection{\textit{Asymmetric Roles and Power Dynamics}} Another notable unique factor in K-12 classrooms is the existing power dynamics among the two human actors. Teachers are responsible for ensuring student success, meeting expectations of parents, and conducting instruction that aligns with the curriculum. This role expectation requires teachers to prioritize monitoring and strong access to student data (which was well reflected in our findings as teachers emphasized access to data). Students, however, are often the subjects of that monitoring with limited agency  \cite{akgun2022artificial}. This difference gives the stakeholders opposing perspectives; teachers may perceive AI as a supporting tool that aids their instruction, whereas students might see it as a surveillance tool (at least for the monitoring aspect). Specifically, in our data, we found that students were afraid of constant AI monitoring and data sharing, which adds unnecessary pressure in their school lives. Teachers, on the other hand, emphasized the need for accessing student data (while acknowledging concerns).

We are aware that there exist similar power dynamics in higher education to some extent \cite{adnin2025examining, luo2025does}. For example, Luo \cite{luo2025does} reports an imbalanced transparency practice in disclosing the use of Generative AI between students and instructors, making students distrust their teachers. Extending this recent line of work, we provide an additional insight from the K-12 education context. Specifically, in our empirical data, we found that teacher-student relationship and trust could be leveraged to alleviate the power dynamics regarding AI use. Participants reported that, as they build trust, they would be able to openly communicate expectations and requirements. While such a mitigation strategy has been proposed in the past to fill the misalignments \cite{luo2025does}, our investigation discovered the importance of the role that trust-building plays in the communication processes.

Lastly, this unique power dynamics is dissimilar to the interactions between parents and children at home (another area within HCI where human-AI interactions have been explored with minors) \cite{10.1145/3687035, zastudil2023generative}, where the relationship is private, with less fear toward judgement (that K-12 students would feel in the classroom, e.g., by peers).

\subsection{Design Implications }
As stated earlier, the gaps in the views would better be aligned so that the AI-supported classroom interactions can be effective and engaging, while considering both stakeholder views. Drawing on the points made in the previous section, we argue how the misalignments could be addressed through design.

\subsubsection{\textit{Addressing Developmental and Metacognitive Differences}} As mentioned above, school students’ limited metacognitive skills may have contributed to the differing perspectives among the stakeholders. In addition to enhancing students’ (and teachers’) AI literacy through explicit instruction \cite{zhou2019metacognitive}, we believe that the misalignments could be improved through a more targeted approach focused on metacognitive scaffolding for students. For example, one could design an AI-based learning system that actively promotes reflective thinking (“I recommend you work on this next topic because I think you have mastered the previous topic. Do you agree?”) instead of merely explaining AI’s decisions (which is a more common approach \cite{amershi2019guidelines} and might be more appropriate for e.g., university students who could actively evaluate algorithms \cite{10.1145/3711104}). Through scaffolding school students’ self-regulated learning with AI \cite{10.1145/3649217.3653621}, students would be able to critically evaluate AI functions, thus helping them better understand consequences of different actions that AI/students/teachers make in the classroom. It is worth noting that university students have generally developed stable self-regulation strategies, whereas younger school-aged students are still more flexible in how they approach learning. As a result, interventions that support metacognition may be especially effective for this group.

We also want to highlight that designers may be able to find space for expectation negotiation \textit{outside} of the AI system to address the metacognitive and developmental differences. One straightforward idea, mentioned by both teachers and students in the study, is to support transparent communications and thus trust building between them on their needs and wants. In addition to activities that promote transparency in AI use \cite{adnin2025examining, luo2025does}, a class of students and their teachers could engage in active perspective-taking exercises to better understand what other stakeholders would feel and experience \cite{yadollahi2022children, nagashima2025understanding}. For example, students could be prompted to think: “If I were a teacher, how would I like to use this tool?” so that students can get to familiarize themselves with how teachers would differently view the technology. This activity could take place off-line without using any digital technology (which may be preferred, according to our findings) to enable a low-cost approach. Moving forward, however, there could also be technological tools that facilitates communications between teachers and students. One idea could be to develop a chat system that connects students and teachers. Although such systems and design concepts exist \cite{aleven2022dashboard}, existing ideas tend to focus on optional communication channels for teachers to check in and offer instructional support; rather, future communication systems could offer active scaffolding to build trust, for instance, by facilitating perspective taking \cite{de2023perspective, yadollahi2022children, nagashima2025understanding} or by letting teachers choose to actively reveal their intention (that teachers would input manually) on instructional strategies and moves to students so that students can get a better idea of what their teacher is intending to do at that moment. These communications would scaffold students’ metacognitive understanding of what would the optimal AI use look like for teachers, and also strengthen the student-teacher relationships that our participants emphasized.

\subsubsection{\textit{Addressing Asymmetric Roles and Power Dynamics}} As mentioned earlier, another unique aspect of K-12 classrooms is the asymmetry of stakeholder roles, which give stakeholders different expectations towards AI. Addressing it requires questioning common assumptions (or default settings of many AI systems) that teachers should oversee initial settings while students remain passive recipients of the system configurations that teachers/schools decide on. Instead, schools/teachers could consider offering students an opportunity to share their preferences and feelings towards certain ways of using AI in the classroom. Giving students increased agency on such decision-making opportunities would help them realize that they are not just a mere subject of AI monitoring but a small contributor for decision making \cite{vincoli2025multidimensional}. Although prior work with university students and instructors has recommended joint decision-making on AI use policies \cite{adnin2025examining, luo2025does}, granting K-12 students equivalent responsibility in this process may pose risks, given their still-developing metacognitive skills (even when students and teachers can openly discuss expectations). Instead, we suggest that primary decision-makers (e.g., teachers) appropriately scaffold students’ participation in decision-making processes, while still affording them a degree of control, for example, by initially focusing discussions on a single feature of an AI system.

From a design perspective, AI-based systems could also support the aforementioned institutional and teacher decisions by offering, for instance, a negotiation dashboard where students can request how parameters of system, teacher, and learner control can be adjusted. Also, for K-12 students, who may gradually develop an understanding of AI through ongoing use, it would be valuable to provide such shared decision-making opportunities not only at the beginning but also periodically throughout the school year.

It is important to note that we do not intend to argue that the asymmetric roles should be entirely reconsidered; Instead, we advocate for exploring design spaces in which students are empowered to actively contribute to decision-making \cite{vincoli2025multidimensional}, thereby making AI monitoring a subject of open discussion. Conversely, it is equally important for teachers to recognize that, due to existing power dynamics, students may feel pressured or stressed by AI monitoring, even if monitoring provides benefits for teachers. To support this awareness, AI-based learning systems could include features that allow students to provide anonymous feedback when they perceive certain AI actions as incorrect or excessive. This would enable teachers to better understand which AI features students consider problematic.

Importantly, however, our study also revealed that student-teacher trust building can make a significant impact on how students and teachers perceive AI use in the classroom. Therefore, we believe that the degree of misalignments depends on how students and teachers have built their relationships within the classroom, and design of any tools or support should be tailored to match the unique context and social dynamics.

\subsection{Limitations }
We acknowledge several limitations of the study. First, the design and the use of storyboards may have introduced certain biases, elicited particular viewpoints and emotions, and/or directed participants’ thinking into particular situations (e.g., focused on classroom interactions only). In addition, our study focused on Intelligent Tutoring Systems (ITSs) as an example of AI tools. While ITSs are one of the major intelligent technologies widely used in school systems \cite{pane2014effectiveness}, the use of other AI tools (e.g., generative AI) could result in different insights. Also, we collected data within Germany, where AI-based tools have not been fully realized in schools. It is possible that, in the future, AI might be used in rather differently than what we depicted in the storyboards, and students’ and teachers’ views might differ once they experience AI systems in classrooms. Further, as we compensated participants, such an incentive might have influenced what the participants shared with us. Moreover, our student and teacher samples included a diverse group of participants in terms of the grade levels represented/teachers have taught. Two teachers taught only students younger than 12 (our student sample included those 12 and above only). This mismatch and diversity in the samples might have influenced how they view student-AI control in the classroom and the resulting themes we have elicited. Also, older students generally tended to share more thoughts than younger students in the study, which might have masked some important, unshared details that only younger students had thought about. Finally, as the data collection happened in Germany with a small sample, our findings would not represent a larger population. We encourage future studies to conduct similar studies with diverse participants, with different cultural contexts to capture rich insights 

\section{Conclusion}

This study investigated how teachers’ and students’ views on AI in the classroom might align/misalign with each other. We found that, despite some aligned views, they have different preferences for AI use in the classroom. In particular, teachers tend to care more about the entire classroom environment while students focus on individual, their own outcomes. Importantly, both students and teachers emphasize the importance of student-teacher relationships that influence how they trust and would use AI systems in the classroom. This work, by conducting systematic, fine-grained investigation and analysis on stakeholder views, makes contributions by offering the knowledge of K-12 students’ and teachers’ (mis)aligned views on AI where human-human relationships have a significant impact on how they perceive and would like to interact with AI systems. Further, we situate the findings within the lens of sociotechnical tradeoffs, highlighting the importance of understanding contexts of AI use. Future research could build on the design implications to help mitigate misalignments among stakeholders and promote more effective and engaging uses of AI in K-12 classrooms.

%%
%% The acknowledgments section is defined using the "acks" environment
%% (and NOT an unnumbered section). This ensures the proper
%% identification of the section in the article metadata, and the
%% consistent spelling of the heading.
\begin{acks}
This work was supported by JST, PRESTO Grant Number JPMJPR2318, Japan. We thank all the participants in the study.
\end{acks}

%%
%% The next two lines define the bibliography style to be used, and
%% the bibliography file.
\bibliographystyle{ACM-Reference-Format}
\bibliography{refs}

%%
%% If your work has an appendix, this is the place to put it.
\appendix

\end{document}